\documentclass{aa}  

\usepackage{graphicx}
\usepackage{txfonts}
\begin{document}

   \title{Observations and NLTE modeling of Ellerman bombs}

   \author{A. Berlicki
          \inst{1,2}
          \and
          P. Heinzel\inst{1}}

   \institute{Astronomical Institute, Academy of Sciences of the Czech Republic,
        25165 Ond\v{r}ejov, Czech Republic\\
              \email{arkadiusz.berlicki@asu.cas.cz}
         \and
             Astronomical Institute, University of Wroc{\l}aw, 51-622 Wroc{\l}aw, Poland,\\}

   \date{Received; accepted}

  \abstract
{Ellerman bombs (EBs) are short-lived, compact,
and spatially well localized emission structures that are observed well in the wings 
of the hydrogen H$\alpha$ line. EBs are also observed in the chromospheric 
CaII lines and in UV continua as bright points located within active regions.
H$\alpha$ line profiles of EBs show a deep absorption at the line center and 
enhanced emission in the line wings with maxima around $\pm$1~{\AA} from the line 
center.  Similar shapes of the line profiles are observed for the CaII IR line
at 8542~{\AA}. In CaII H and K lines the emission peaks are much 
stronger, and EBs emission is also enhanced in the line center.}
 {It is generally accepted that EBs may be considered as compact microflares 
located in lower solar atmosphere that contribute to the heating of these 
low-lying regions, close to the temperature minimum of the atmosphere. 
However, it is still not clear where exactly the emission of EBs is formed in the 
solar atmosphere. High-resolution spectrophotometric observations of EBs 
were used for determining of their physical parameters and construction 
of semi-empirical models. Obtained models allow us to determine the position of 
EBs in the solar atmosphere, as well as the vertical structure of the activated EB atmosphere}
 {In our analysis we used observations of EBs obtained in the H$\alpha$ and CaII H 
lines with the Dutch Open Telescope (DOT).  
These one--hour long simultaneous sequences obtained with high temporal 
and spatial resolution were used to determine of the line emissions.
To analyze them, we used NLTE numerical codes for the construction of grids of 243 
semi--empirical models simulating EBs structures. In this way, the observed 
emission could be compared with the synthetic line spectra calculated for all such
models.}
 {For a specific model we found reasonable agreement between the observed and 
theoretical emission and thus we consider such model as a good approximation 
to EBs atmospheres. This model is characterized by an enhanced temperature 
in the lower chromosphere and can be considered as a compact 
structure (hot spot), which is responsible for the emission observed in the wings 
of chromospheric lines, in particular in the H$\alpha$ and CaII~H lines. }
   {For the first time the set of two lines H$\alpha$ and CaII~H was used to construct 
semi--empirical models of EBs. Our analysis shows that EBs can be described by 
a "hot spot" model, with the temperature and/or density increase through 
a few hundred km atmospheric structure. We confirmed that EBs are located close to the 
temperature minimum or in the lower chromosphere. Two spectral features 
(lines in our case), observed simultaneously, significantly strengthen the constraints 
on a realistic model.}

   \keywords{Sun: chromosphere -- Line: profiles -- Techniques: imaging spectroscopy -- Radiative transfer}

   \maketitle
%

\section{Introduction}

Ellerman bombs (EBs) or moustaches were first described by Ellerman (1917) 
as small-scale (approx. 1{$\arcsec$} or less) structures observed in the wings of the 
hydrogen H$\alpha$ line.
Ellerman himself has described them as bright, compact structures of a few 
minutes lifetime, which can by seen in spectrograms taken in the 
range from 4 - 15{\AA} from the center of the H$\alpha$ line.

EBs are observed in regions of the magnetic flux emergence or 
at the ends of chromospheric fibrils forming superpenumbra around large sunspots. They occur not only in a new 
active region but also in old ones, where the magnetic field starts to rebuild.
Georgoulis et al. (2002) found that EBs have a tendency to concentrate say, above the magnetic neutral 
lines outlining the boundaries of supergranular cells or in MDF (moving dipolar 
features, Bernasconi et al. 2002).

The average lifetime of EBs was analyzed by  Severny (1956), McMath et al. (1960), Roy \& Leparskas (1973),
and Kurokawa et al. (1982), who found that the average value is around a dozen 
minutes, but there are also observations of EBs lasting for more than half an 
hour (Roy \& Leparskas 1973, Kurokawa et al. 1982, Qiu et al. 2000, Pariat et al. 2007).
Quite recently, Nelson et al. (2013) presented an extended statistics on the lifetimes and
sizes of EBs.
The H$\alpha$ light curves of those EBs, which were observed for 
more than 20 minutes,  seem to consist of several maxima or 
show one flat maximum with some oscillations of brightness (Roy \& Leparskas 1973, 
Kurokawa et al. 1982, Pariat et al. 2007).

Except for the hydrogen H$\alpha$ line, EBs are also observed in CaII~H, K, and
infrared 8542~{\AA} chromospheric lines. Such observations are presented, 
e.g., in Pariat et al. (2007), Yang et al. (2013) or Vissers et al. (2013), 
where the CaII~8542~{\AA} line profiles emitted from EBs have a similar shape 
(dark line center, enhanced line wings) as well known H$\alpha$ profiles. 
There are very few observations of EBs in CaII~H and K lines, but the old spectra 
obtained by Kitai \& Kawaguchi (1975) show that the EBs emission 
in CaII~K line is strongly enhanced.

The line profiles and continua observed in EBs have been used in the past
in a few works to construct models of the atmosphere within these 
structures. The results of this modeling imply that 
the whole region of EBs is hotter than a quiescent solar atmosphere, 
and the heating is significant only in the lower 
chromosphere or even photosphere (see review by Rutten et al. 2013), 
where the H$\alpha$ line wings and UV continuum are formed.
Fang et al. (2006) have obtained a semiempirical model
of EBs, and they find that EBs require extra 
heating in the lower atmosphere, with about a 600-1300\,K temperature enhancement 
close to the temperature minimum region.
They also calculated that the total energy of EBs is about 
$10^{26}$ to $5\times10^{27}$\,erg, and they suggest 
that EBs could be similar to microflare events. However, they show a significant
temperature increase even in regions of the middle and upper chromosphere, which
should lead to a substantial increase in the H$\alpha$ line-center intensity. 
This increase in the H$\alpha$ line-center intensity is not observed in EBs
unless a strong absorption in the overlying canopy is considered (Rutten et al. 2013). 
We comment on this problem in the discussion section.

Kitai (1983) performed NLTE calculations and derived theoretical H$\alpha$
line profiles emitted by different models of EBs. The models are characterized by 
a temperature and density increase located in the middle chromosphere. He was able to
explain the basic characteristics of the spectra obtained at Hida Observatory. For recent 
reviews on EBs see Kitai (2012) and Rutten et al. (2013).

The principal aim of this paper is 
to find a model that can describe the structure of EBs in the solar atmosphere 
and for which the theoretical emission in spectral lines will reasonably reproduce 
the observed one. Using the NLTE (departures from local thermodynamic equilibrium)
numerical calculations, we simulated the observed emission and found a 
satisfactory model of EB. For the first time, the simultaneous observations in the
two lines H$\alpha$ and CaII H were used to compare them with the 
synthetic spectra calculated with the NLTE codes.
We used the observations obtained with Dutch Solar Telescope (DOT, Rutten et al. 2004)
in different band-passes. The DOT telescope does not provide the 
spectroscopic data, but the filtergrams obtained in different lines allow 
us to compare them with the synthetic spectra at various wavelength positions.

\section{Observations of Ellerman bombs and their multiwavelength emission }

   \begin{figure*}
          \centering
            \includegraphics[width=15cm]{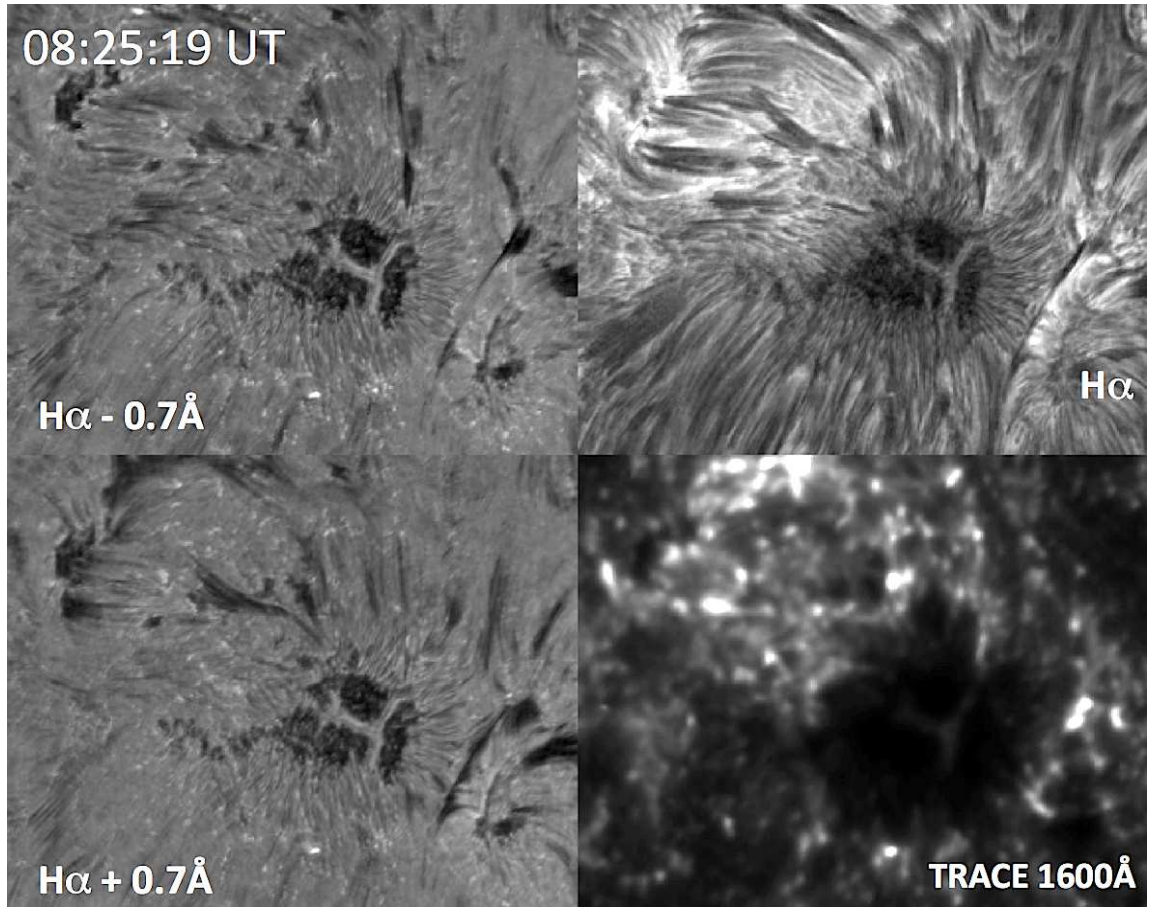}
      \caption{Example of Dutch Open Telescope observations of the active region
      NOAA 10892 on June 7, 2006 at 08:58 UT in the H$\alpha$ line and in TRACE 1600~{\AA}.}
         \label{ar_dot}
   \end{figure*}

In the present analysis we use the observations 
of EBs obtained in two lines (H$\alpha$ and CaII~H) which can be modeled
with our NLTE codes. Since EBs are compact structures  (diameter less than 800~km),
to avoid the contamination of their emission from the nearby 
solar atmosphere, it is also necessary to use high-resolution observations 
obtained under good seeing conditions. The EBs emission obtained 
from low resolution observations can be affected by the radiation 
of other structures around.

Therefore, we inspected the database of the Dutch Open Telescope (DOT, Rutten at al. 2004)
and found a series of high-quality observations obtained in the H$\alpha$ and CaII~H 
lines, as well as in the G-band spectral range. These observations were obtained on June 7, 
2006 and concern the 
active region NOAA 10892 (Fig. 1). This active region was located close to 
the solar disk center, so the emission of EBs was not affected much by surrounding 
chromospheric structures. It would be difficult to compare the 
observed radiation of EBs contaminated by some neighboring structures with the 
theoretical emission calculated without including any other emitting structures.

On June 7, 2006, active region 10892
consisted of a big leading sunspot that had the negative magnetic field polarity and of a group 
of smaller sunspots with the positive polarity.
This active region was also simultaneously observed by TRACE (Transition Region 
and Coronal Explorer) satellite in the ultraviolet 1600~{\AA} channel (Fig. 1).

The DOT observations were  obtained between 08:20 and 09:30 UT with a cadence of 
around 30 seconds. The DOT data consists of 134 images in each separate bandpass 
around the H$\alpha$ line ($-0.7$~{\AA}, line center, $+0.7$~{\AA}), in the CaII~H line 
center and wing at $-2.35$~{\AA}, as well as in the G-band. The TRACE sequence consists 
of 146 images, but some of them (between 09:00-09:15 UT) were not useful owing to 
calibration problems.

In total, we used around 950 image arrays to determine the time evolution of EBs
intensities in different spectral windows. To make precise brightness determination,  
all images in all bandpasses were coaligned and saved in 
one data cube. Since ground-based observations were affected by the atmospheric scintillation,
the accuracy of the coalignment is around 1{$\arcsec$}. Using our data cube, we could then 
plot the time evolution of emission of any structure that was visible within the field-of-view (FOV)
of the DOT telescope.

   \begin{figure*}
   \centering
   \mbox{
            \includegraphics[width=9cm]{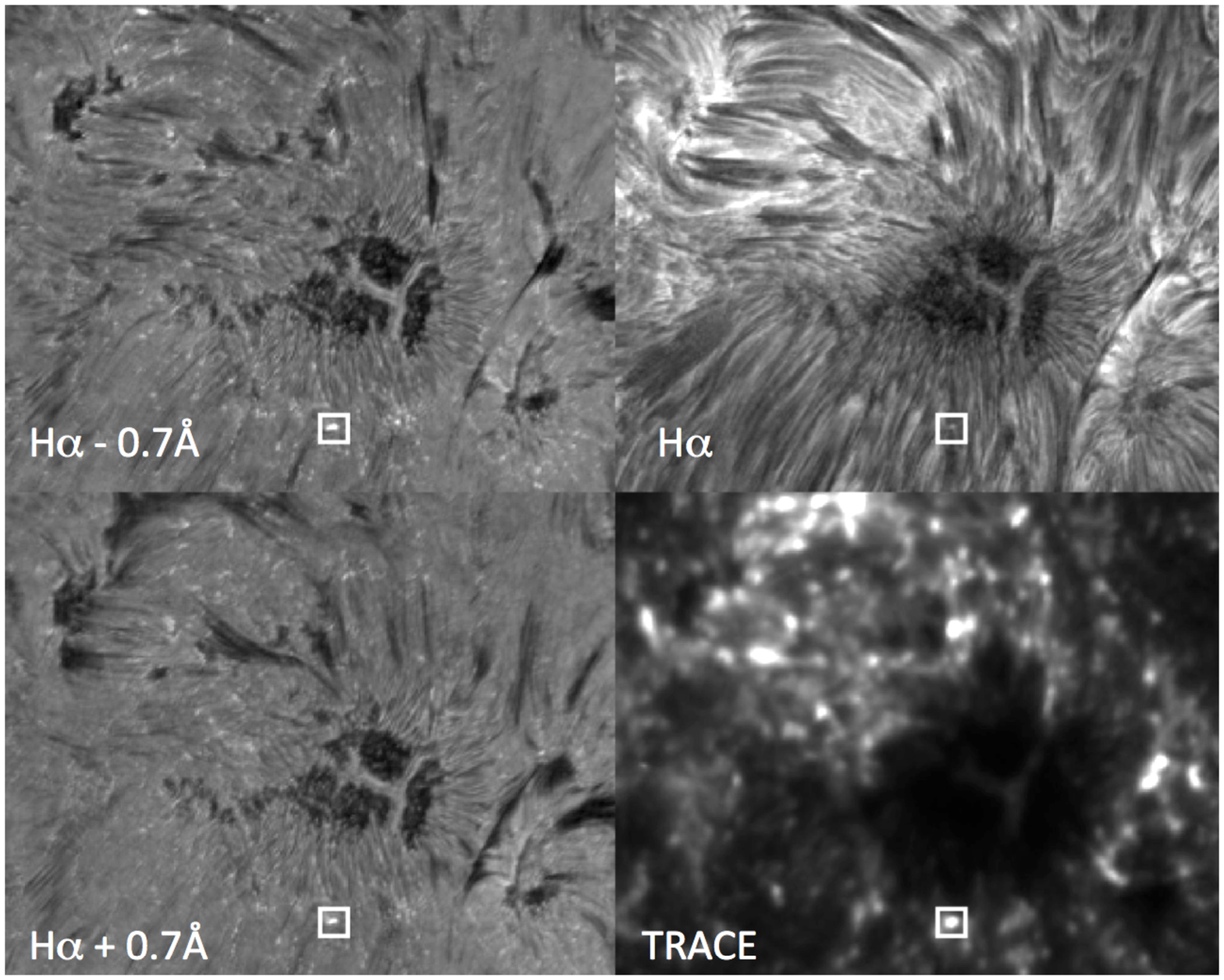}
            \includegraphics[width=9cm]{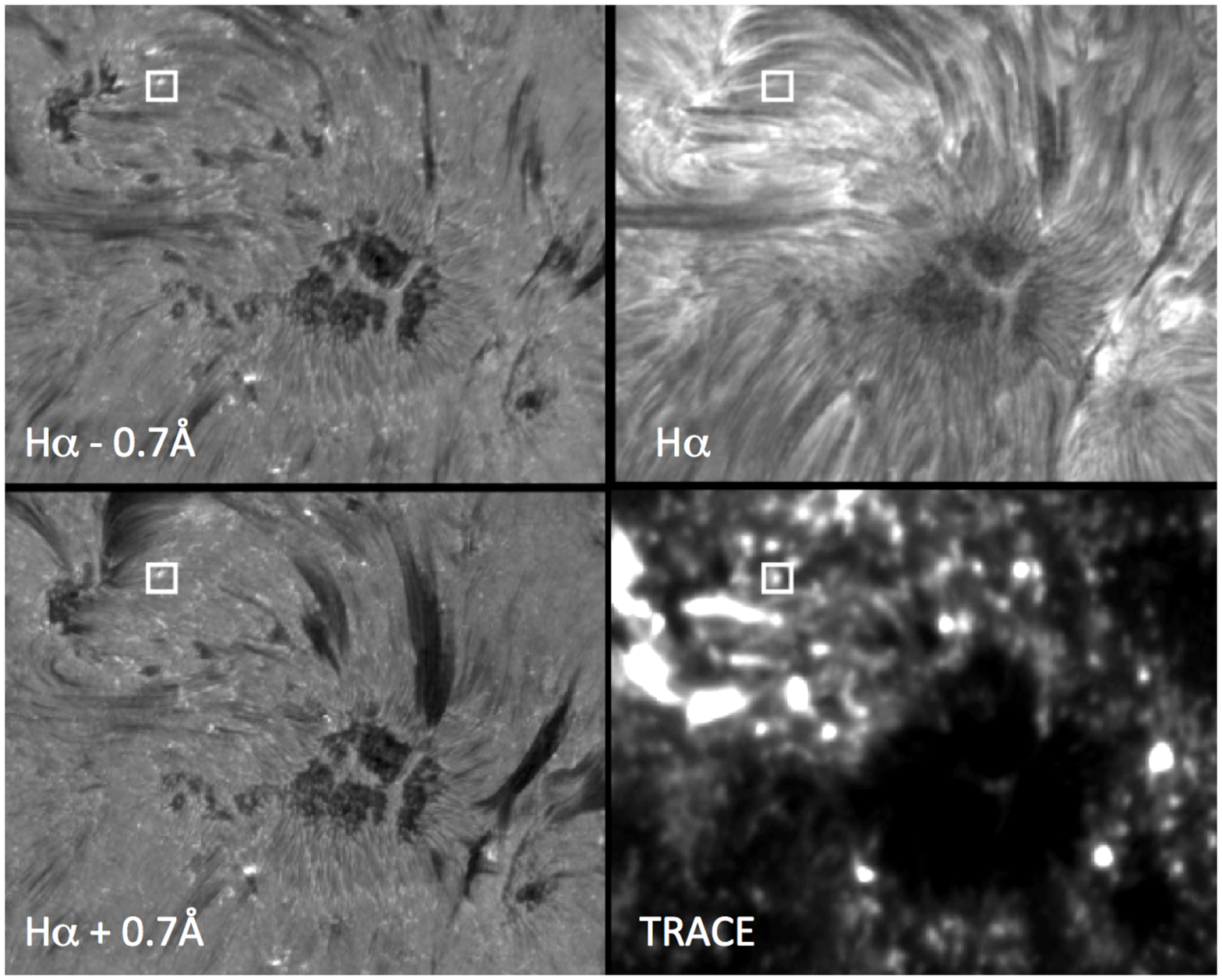}
            }
      \caption{DOT and TRACE images of the active region with the Ellerman bomb EB\_1(left, 08:25 UT) and EB\_2 
      (right, 08:53 UT). EBs are marked with the white boxes.}
         \label{ebs_box}
   \end{figure*}

In the next step we carefully analyzed time sequences of the data obtained in the H$\alpha$-$0.7$~{\AA}.
According to previous analyses, EBs should be visible in this wavelength thanks to their enhanced emission 
in the line wings. After inspection of the data cube, we found a few brightenings visible in H$\alpha$-$0.7$~{\AA} 
at different times during the observations. In the contrast, in observations obtained in the H$\alpha$ 
line center we did not find any brightenings corresponding to those observed at -$0.7$~{\AA}, which 
agrees with previous observations. In Figure 2 we present filtergrams with two 
different EBs observed at different times. EB, which was observed at 08:25 UT, was denoted as EB\_1,
the one observed at 08:53 UT - EB\_2. One may notice that EBs are almost not visible in the image 
taken in the H$\alpha$ line center.  

It is important to notice that for all structures within 
the FOV, the DOT data does not provide us with the intensity in one specific wavelength,  
but we have the intensities integrated over the passbands of the DOT filters. Therefore,
for further analysis and comparison of the observed and theoretical emissions, it is important 
to know the parameters of such filters.

Moreover, to avoid problems with the absolute photometric calibration, 
affected by varying observing conditions and uncertainties connected with DOT filters bandpasses, 
we used the contrast $C$ parameter instead of the intensity. The contrast is defined as:

\mathindent1.0cm
\begin{equation}
C = (I_{EB} - I_{QS})/I_{QS} \, ,
\end{equation}

where $I_{EB}$ is the intensity of EB observed at a given moment of time, 
and $I_{QS}$ is the intensity of radiation emitted by the quiet-Sun atmosphere. Both values of 
$I_{EB}$ and $I_{QS}$ were derived from the DOT filtergrams.
The intensity $I$ as detected by DOT is the {\em integrated intensity}, i.e. the specific
intensity emitted by an EB, multiplied by the transmittance profile of the DOT filter at a given
wavelength position and integrated over the wavelength range of the filter transmittance and
$I_{QS}$ was determined 
by averaging part of the DOT images, located far from the big sunspot, without any active 
structures around. Such quiet-Sun areas can be, for example, recognized in the lower lefthand corner 
of the DOT images (Fig. 1).  Both the emission 
of EBs and the emission of the quiet-Sun can be affected during observations by changeable 
weather and by atmospheric and instrumental conditions. However, the contrast $C$ as defined above 
is not sensitive to these influences, allowing us to perform the analysis of time variations of EBs emission.

In Figure 3 we present the time evolution of the emission for EB\_1 (left) and EB\_2 (right) in different 
spectral ranges. The vertical axes represent the contrast defined in Eq. 1.
The time changes of the contrast emission presented in Fig. 3 clearly show that EBs 
are visible mainly in the H$\alpha$ line wings and in CaII~H line center. EBs are also 
very bright in TRACE 1600~{\AA} channel, but their emission in the calcium line wing and in G-band
is less enhanced, while EBs are almost not visible in the H$\alpha$ line center. For further detailed analysis, 
we used the contrast observed in the H$\alpha$ and CaII~H lines at the time of maximum 
brightness of EBs. For EB\_1 the maximum contrast in H$\alpha$ line wings is around 1.6, and the contrast 
in the line center is 0.3. For Ca H the observed contrast in the line wing at $-2.35$~{\AA} is 1.5, 
and in the line center is 2.6. EB\_2 is characterized by lower brightness and lower contrast . These 
observed contrast values will be used for comparison with our theoretical calculations.

It is interesting to notice 
that for both EBs the maximum contrast in the red and blue wings of the in H$\alpha$ line is similar.
Small differences are probably due to the asymmetries of line profiles probably caused 
by the slow vertical plasma flow in EBs. The low contrast observed in the H$\alpha$ line center 
is consistent with previous results and clearly shows why EBs are not visible in the line core.
From these plots it is also possible to determine the approximate 
lifetime of EBs: around 15 minutes for EB\_1 and 12 minutes for EB\_2.

In the next part of the paper we compare the observed contrast of EBs with theoretical results 
obtained from the NLTE modeling.

   \begin{figure*}
   \centering
   \mbox{
            \includegraphics[width=9cm]{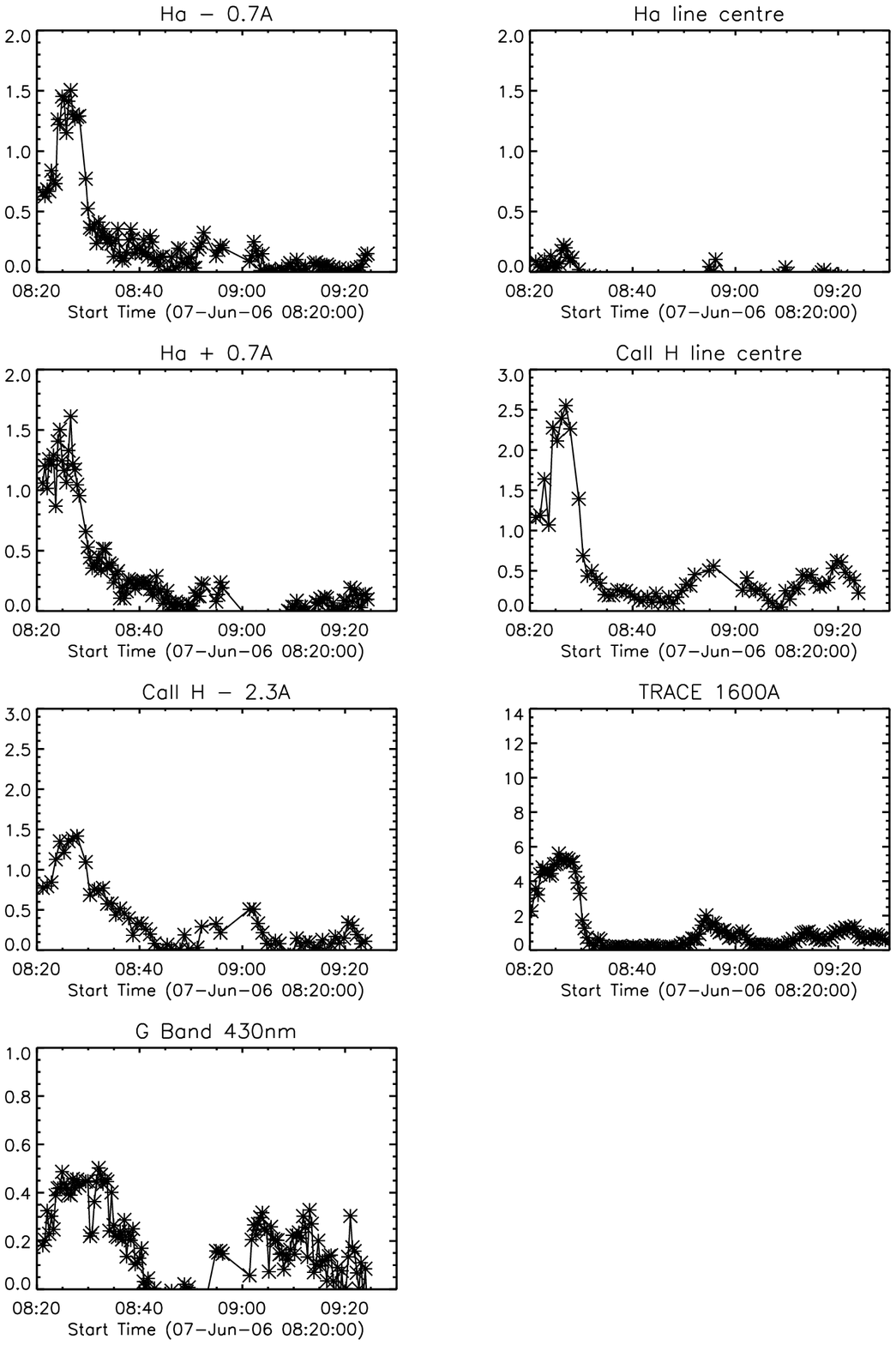}
            \includegraphics[width=9cm]{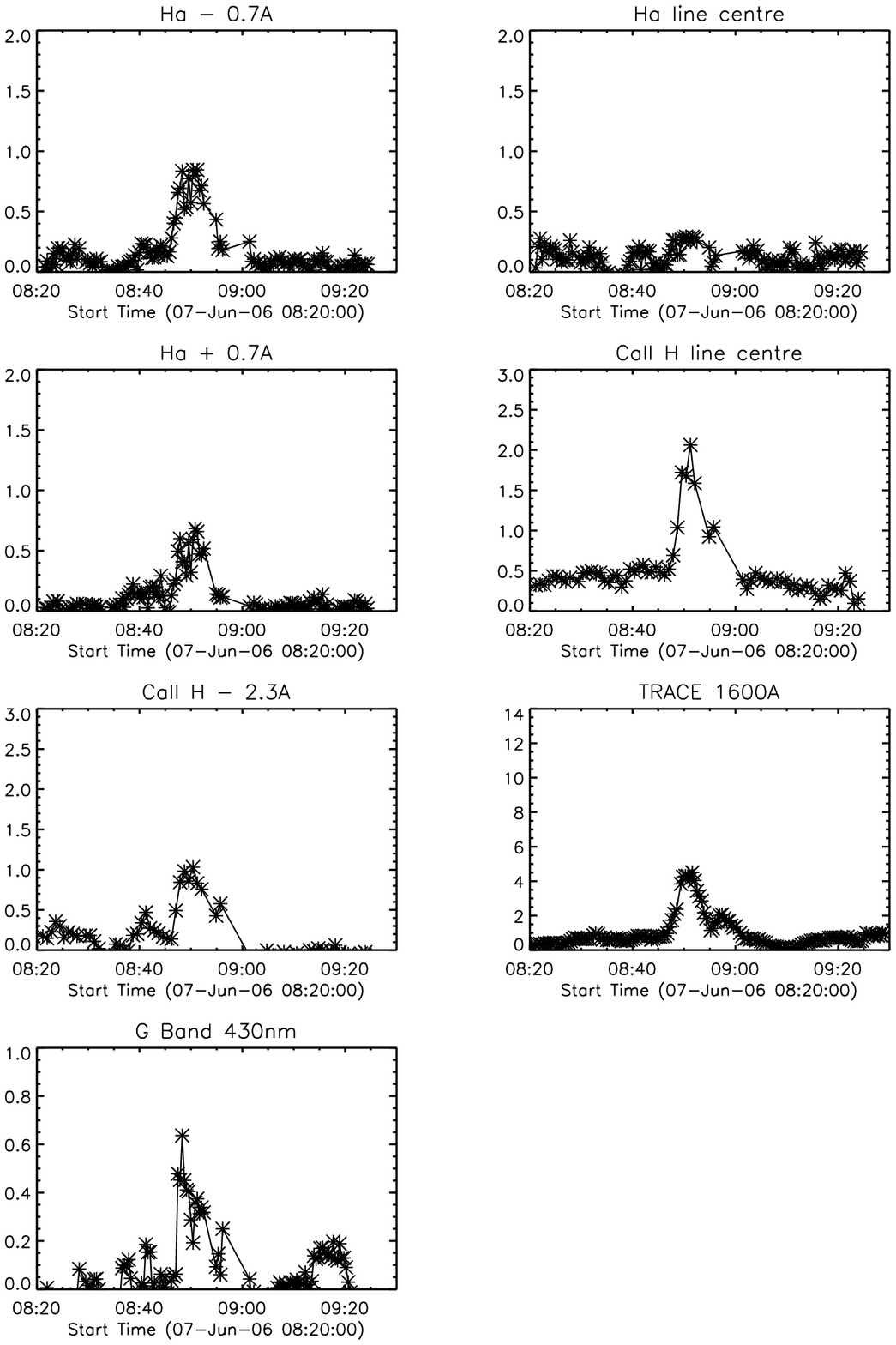}
            }
      \caption{Time evolution of the emission of EBs No. 1 (EB\_1, two leftmost columns) and 
                      No. 2 (EB\_2, two rightmost columns) in the H$\alpha$, 
                     CaII~H, and G-band spectral ranges and in TRACE 1600~{\AA} channel. The vertical axes represent the 
                     contrast defined in Eq. 1. }
         \label{ebs_time}
   \end{figure*}

\section{NLTE modeling of EBs}

\subsection{Methods}

Realistic synthesis of the EB's spectra is based on the multilevel NLTE radiative-transfer
modeling. However, two important aspects have to be considered: the atmospheric model and its geometry.
So far, all transfer models of EBs have been based on a simple 1D plane-parallel semi-infinite atmosphere, assuming
that the horizontal extension of EBs is larger than or at least comparable to their height extensions (see, e.g., 
Kitai (1983), Fang et al. (2006), Socas-Navarro et al. (2006), Berlicki et al. (2010)). As mentioned in the 
Introduction, EBs have typical sizes on the order of about 1{$\arcsec$}, while the range of heights where 
their optical spectra are formed is somewhat smaller (a few hundred km). The 1D models thus represent a reasonable
first approximation, they can reproduce the spectra in central parts of the observed structures. 
Moreover, as we see later, the most realistic models from our extensive grids are located at the base
of the chromosphere and some may extend down to the region of the temperature minimum. At these heights, the atmospheric
conditions together with an enhanced temperature of EBs (the so-called hot spot model), lead to departures from LTE
smaller than in the middle or upper chromosphere. Under strict LTE, the geometry problem is irrelevant
because the line synthesis reduces to the formal solution along the line-of-sight (LOS) with the Planck source function.
Since we do not intend to study the spatial variations in EBs intensities and since our spectral resolution is
rather poor, we use in this study the 1D models. Such models are also very convenient for the construction
of large grids of models, which is our aim here. 

Another problem with the NLTE modeling is how to construct the realistic atmospheric model of the EB.
This issue seems to be very problematic and so far only a little attention has been devoted to it. Standard semi-empirical
models of the solar atmosphere assume the hydrostatic equilibrium and any adjustment of the kinetic temperature
at certain heights affects the new equilibrium state. For EBs, this approach has recently been used by 
Socas-Navarro (2006) and Fang et al. (2006).
The latter authors found not only a hot spot structure in the lower chromosphere, 
but also rather significant enhancement of temperature in the middle and
upper chromospheres. On the other hand, Kitai (1983) in his exploratory NLTE models of EBs enhanced the
kinetic temperature and/or gas density locally (as a box-like enhancement on a prescribed height scale) and
left his models out of hydrostatic equilibrium. 

There are MHD models of the formation of EBs (e.g., Archontis et
al. 2009), but they consider a fully ionized plasma and do not provide atmospheric models that could be
directly implemented in the NLTE codes. Therefore, we use here a similar approach to that of Kitai (1983) and
modify the temperature and gas density at prescribed range of heights - we call this a hot spot model in analogy to flare
modeling on dMe stars (Kowalski et al. 2013). We start with the semi-empirical model of the quiet solar
atmosphere C7 published recently by Avrett \& Loeser (2008) and enhance the temperature within
the hot spot region in the form of a Gaussian hump. Maximum temperature enhancement, together with the 
central position of the hot spot, 
represent free parameters of our models, in addition to a density enhancement that we take simply as a constant within
the hot spot region. Therefore, each model has four free parameters: position of the hot spot, its width, peak temperature
and density enhancement. 

Our hot spot models are no longer in hydrostatic equilibrium and they are computed on the geometrical scale $z$. 
The problem of EBs atmospheric structure will require much more attention in the future and 2D/3D radiation-MHD simulations will be needed to adequately describe the EBs atmosphere. The situation is even more complex due to
the presence of obscuring chromospheric fibrils, spicules, or jet-like flows associated with EBs 
(Rutten et al. 2013), with velocities in the range of 10 - 20 km 
sec$^{-1}$ and the expected complex structure of the emerging and reconnecting magnetic field 
(e.g., Archontis et al. 2009). 

Based on the atmospheric model described above, we compute the NLTE synthetic spectra of the observed lines
H$\alpha$ and CaII H. First, we solve the NLTE transfer problem for a five-level plus continuum hydrogen model,
using the angle-averaged partial-frequency redistribution (PRD) in the Lyman $\alpha$ and Lyman $\beta$ lines. 
From the converged solution, 
we synthesize the emergent H$\alpha$ line profile of EB. In the second step we use the same model and computed
electron densities to synthesize the CaII spectrum, especially the CaII H line that we have observed. We use a five-level
CaII model atom with atomic data and collisional rates provided by Shine \& Linsky (1974) and H. Uitenbroek (private
communication). The CaII code uses PRD in both resonance lines H and K and
was tested against the results of Uitenbroek (1989). In the hydrogen code, the electron densities are 
modified to account for the metal ionization because it is dominant around the temperature-minimum region. The line
profiles of hydrogen and CaII lines are computed with the Voigt functions, where the damping parameters 
primarily depend on the natural and Stark widths. The Doppler width contains the microturbulent broadening with velocities
taken from the model C7 (for modifications see below). The multilevel NLTE transfer problem is solved using
the multilevel accelerated lambda Iteration technique (MALI) of Rybicki \& Hummer (1991, 1992) and the linearization
scheme to account for the hydrogen ionization equilibrium (Heinzel 1995). 

\subsection{Grid of models}

Using the fast MALI code, we were able to construct a large grid of 1D hot spot models. As stated above, these
models are based on the C7 atmospheric model of Avrett \& Loeser (2008) and the hot spot temperature enhancement
is modeled by a Gaussian-like hump on top of the original temperature structure. Two parameters describe the
temperature: maximum temperature increase and height position of the center of the hot spot. The maximum 
temperature increase varies in the range from 1100 K to 5500 K, with steps of 550 K for different models. The height 
position of the temperature hump is between 100 and 1200 km above the $\tau_{500}$=1 layer in the solar photosphere. 
In successive models we decreased this height by approximately 40~km steps. 
The width of temperature hump is the same for all our models and amounts to 400 km.
In total, we constructed 243 models with 9 different hot spot temperature enhancements in 27 different height locations
in the atmosphere. For all these models, we computed the synthetic H$\alpha$ and CaII~H line profiles up to far wings,
using our MALI codes. We constructed three grids of such models, with the gas-density structure identical to that of C7,
twice high density in the hot spot region and four times higher density. This corresponds to density 
enhancements considered by Kitai (1983). For these models we kept the same microturbulent velocities 
as in C7, but we also added one more grid with C7 density structure, but three times larger microturbulence 
in the hot spot region. In addition, for test purposes, we also computed 
one grid using complete frequency redistribution (CRD) in CaII lines.

   \begin{figure*}
   \centering
   \mbox{
            \includegraphics[width=5cm]{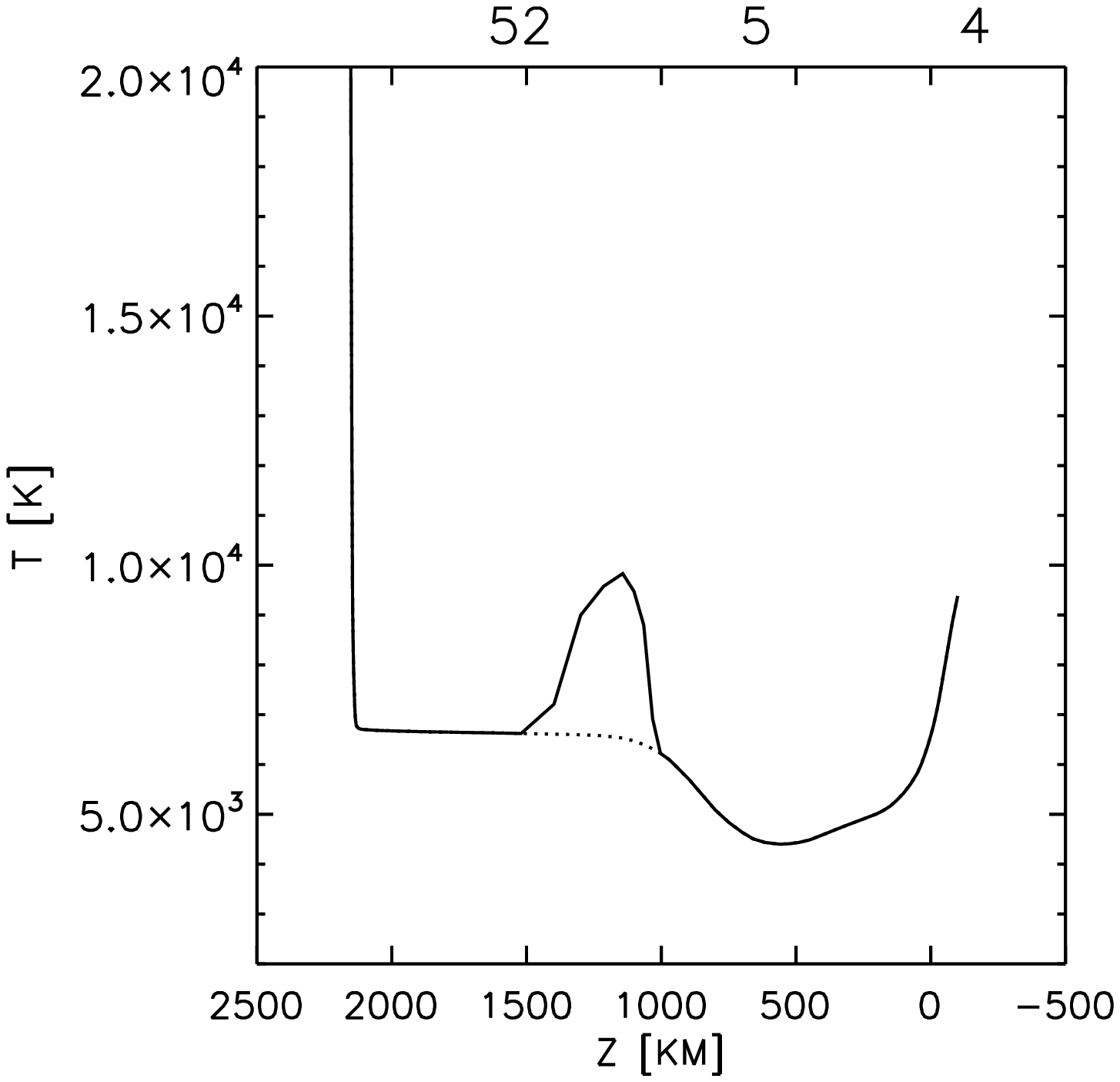}
            \includegraphics[width=5cm]{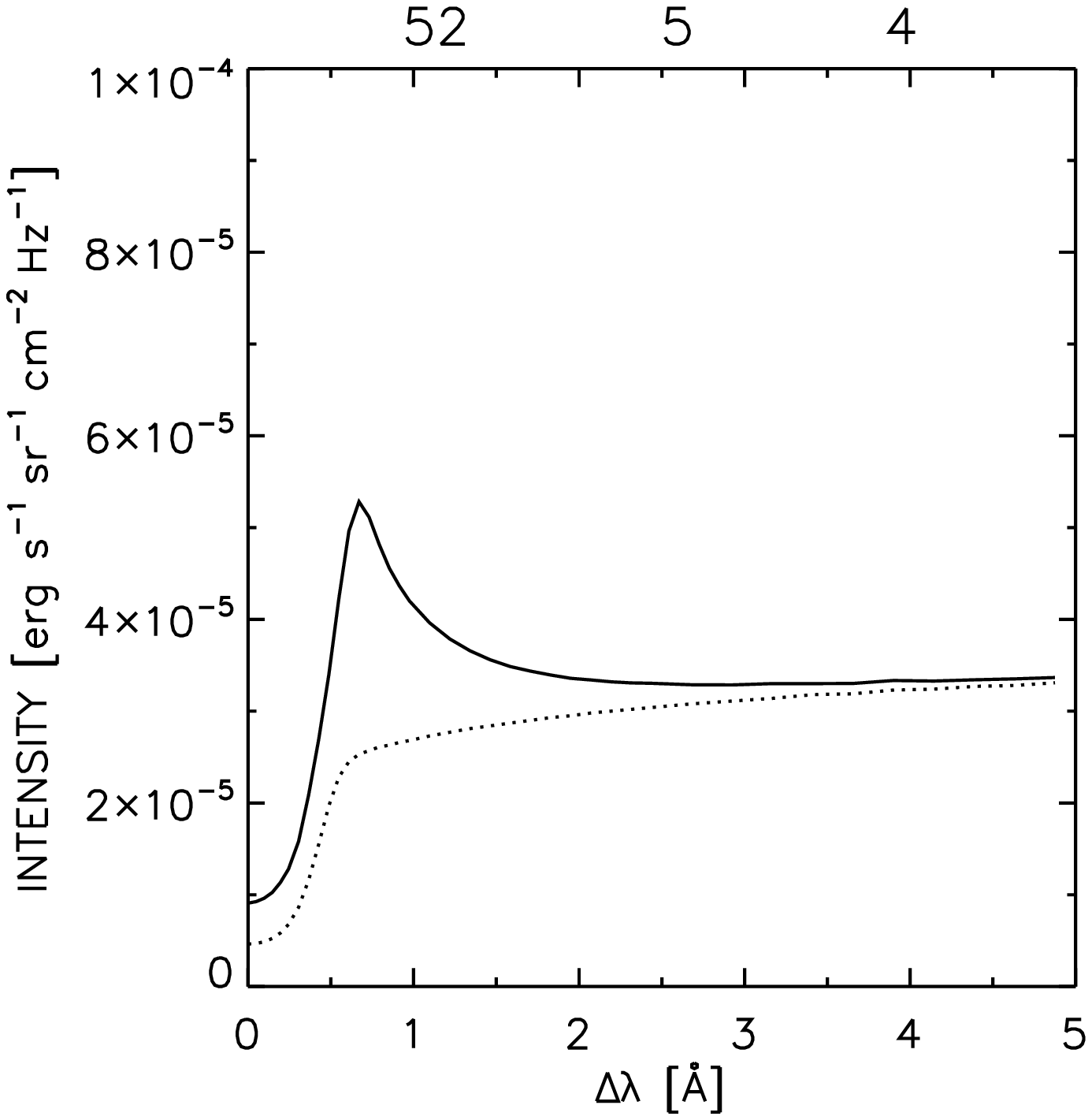}
            \includegraphics[width=5cm]{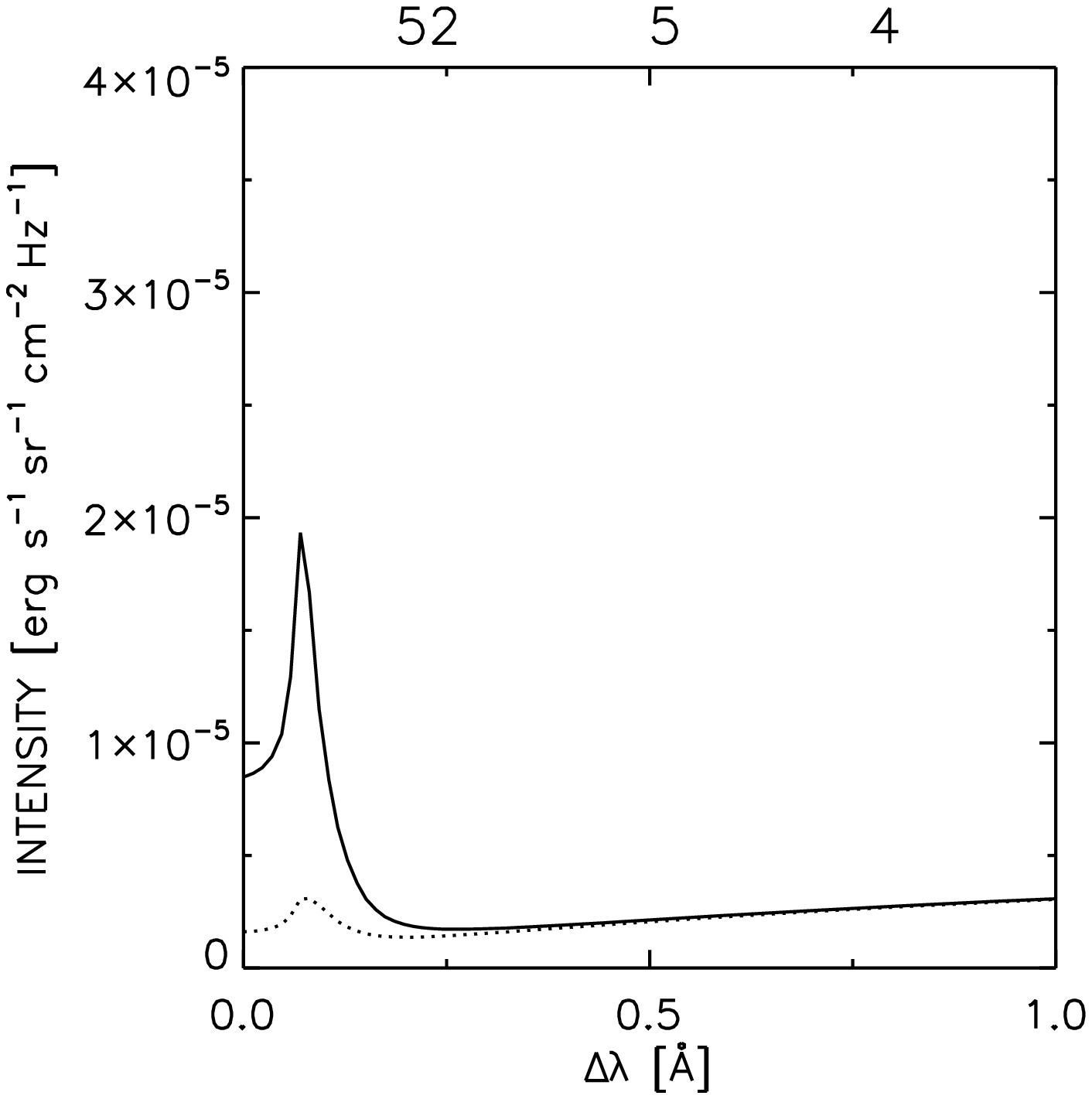}
            }
   \mbox{
            \includegraphics[width=5cm]{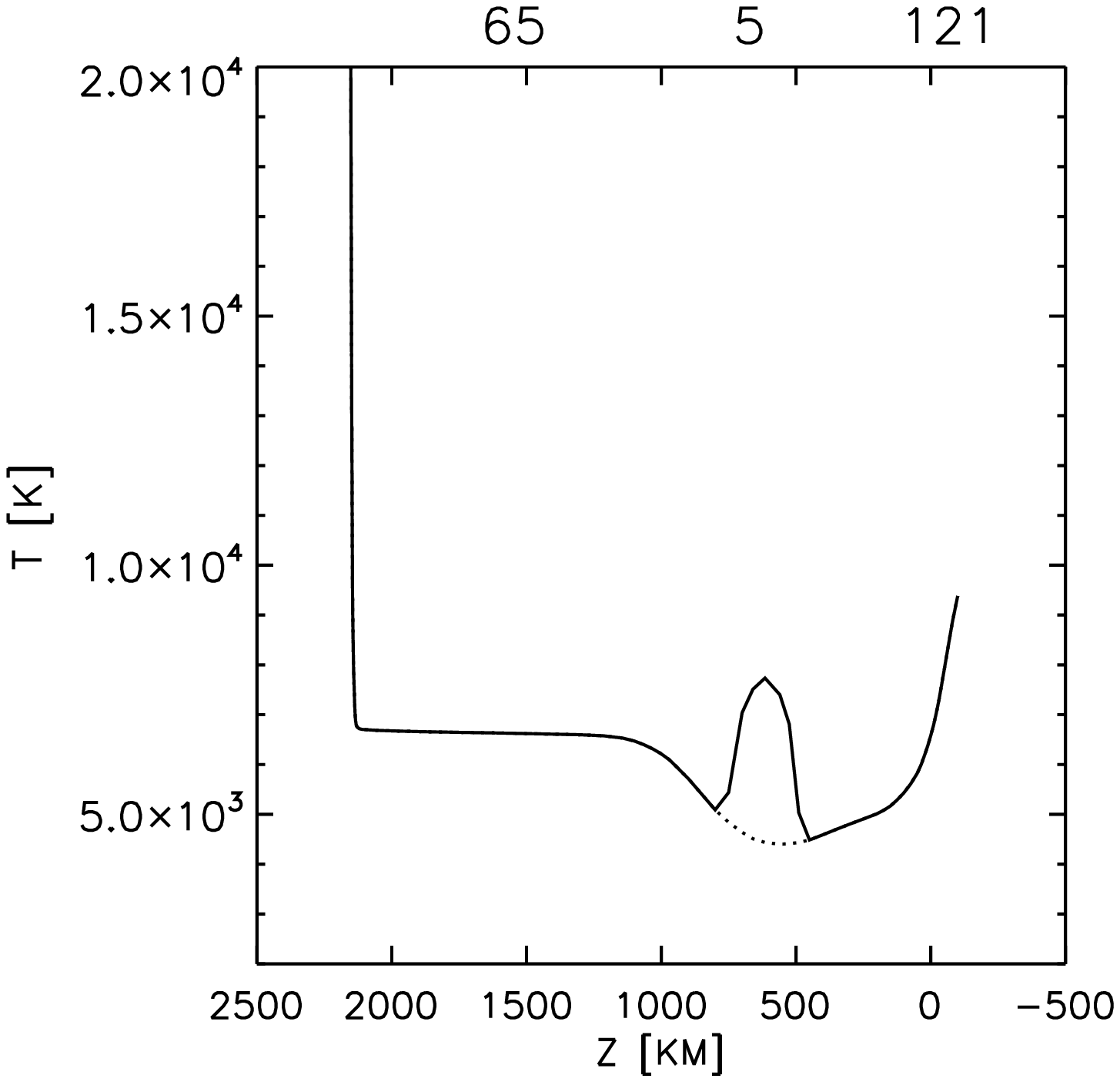}
            \includegraphics[width=5cm]{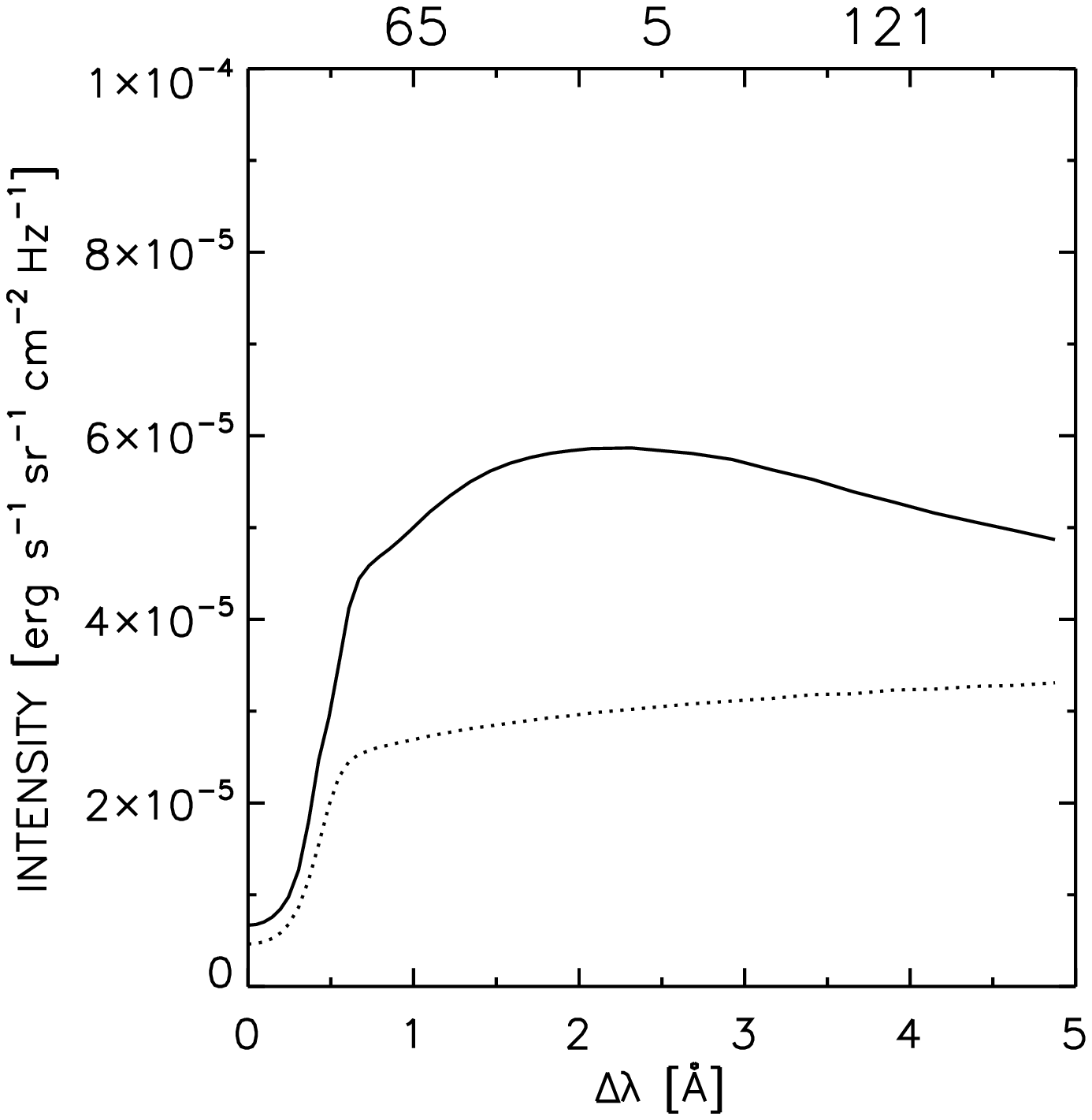}
            \includegraphics[width=5cm]{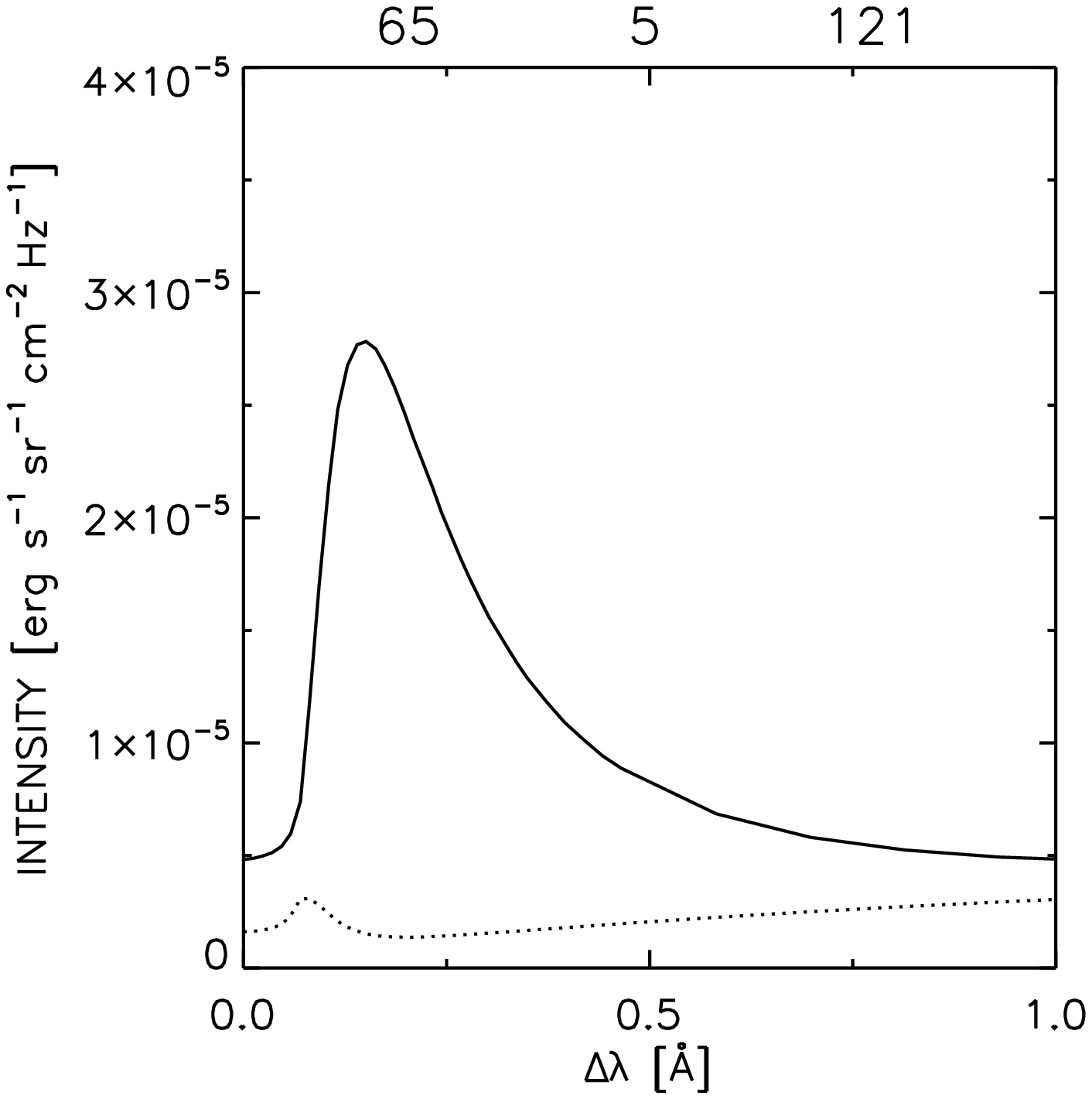}
            }
      \mbox{
            \includegraphics[width=5cm]{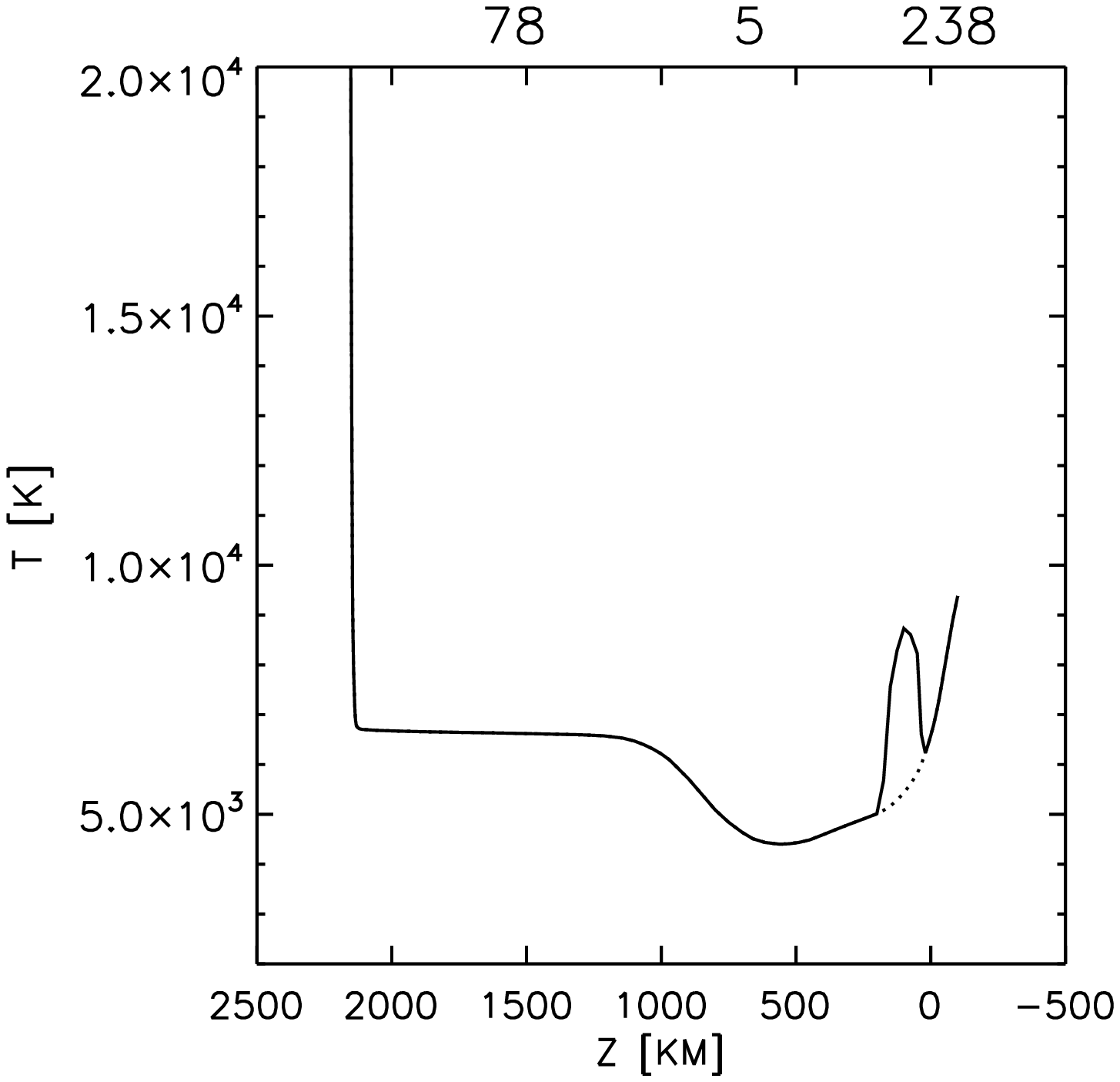}
            \includegraphics[width=5cm]{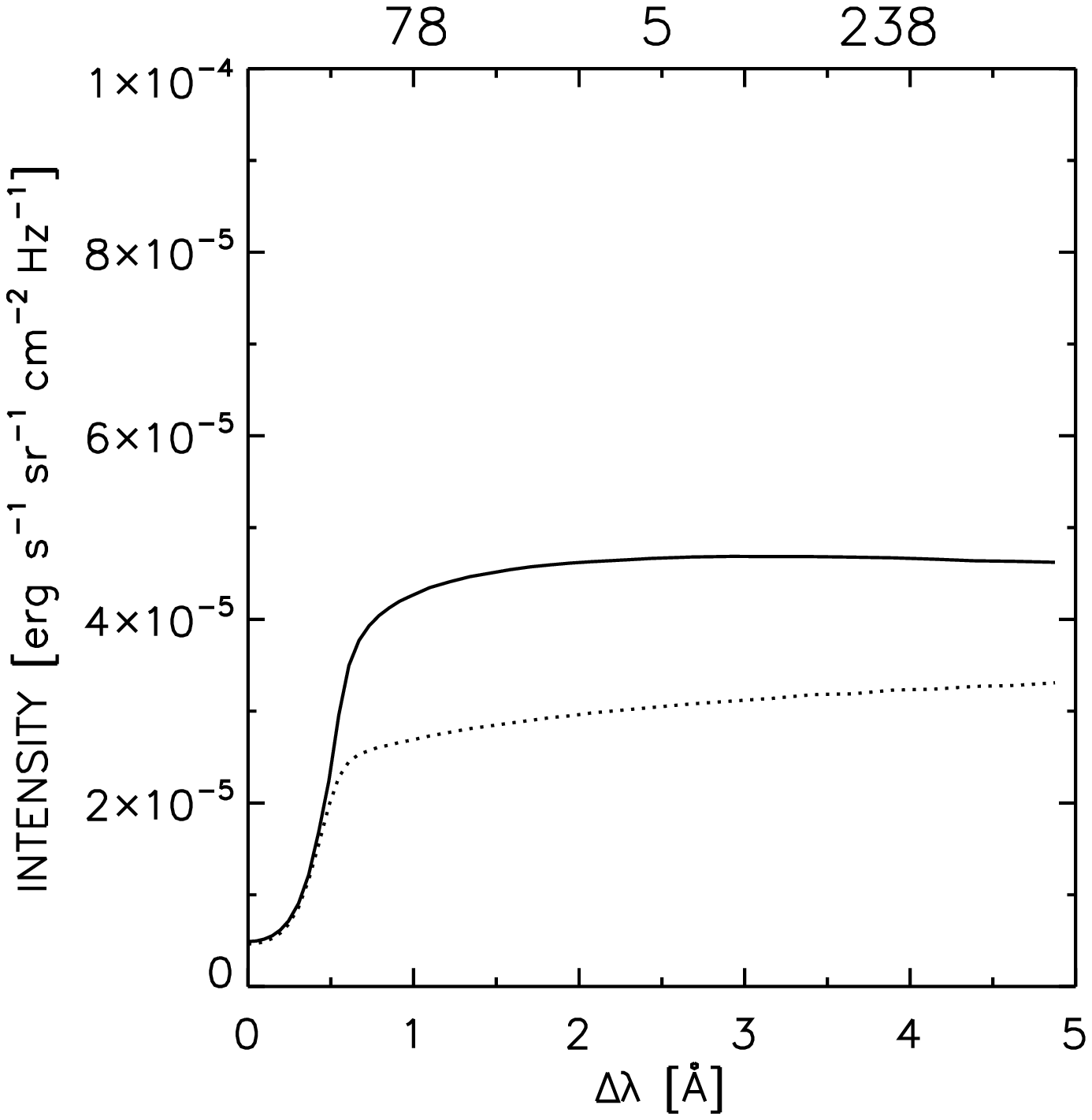}
            \includegraphics[width=5cm]{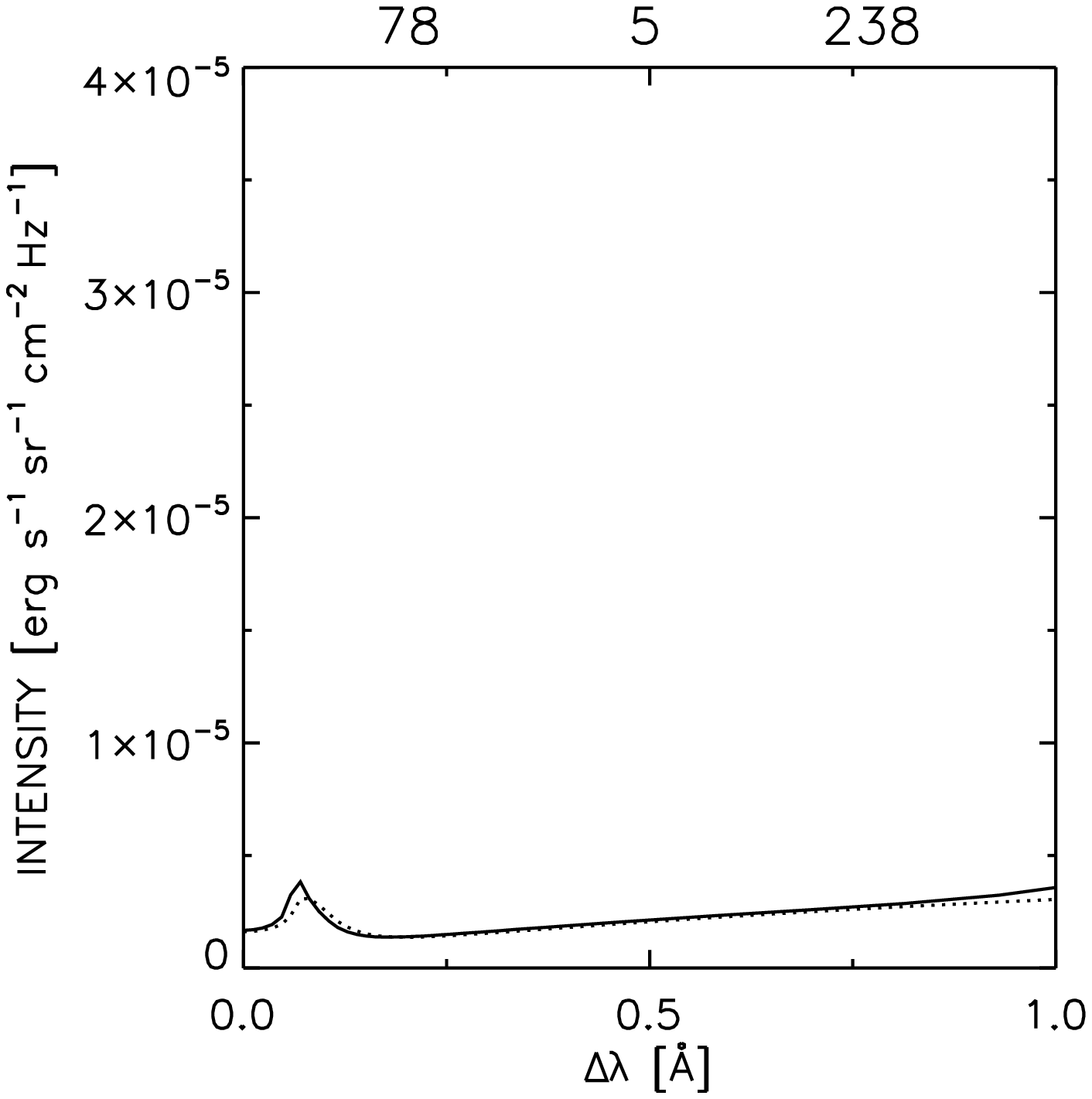}
            }           
      \caption{Example of hot spot models of EBs: temperature distribution (left),
                     synthetic H$\alpha$ line profiles (middle), and CaII~H line profiles (right). Dotted lines 
                     represent the temperature distribution of the C7 model of the quiet-Sun atmosphere 
                     and corresponding line profiles of the quiet Sun. The numbers above the plots 
                     denote the specific model numbers. We chose three representative models (4, 121, and 238
                     in upper, middle, and lower rows) 
                     in the beginning, in the middle, and in the end of the grid with the gas-density structure 
                     identical to that of C7 (see Sect. 4 for details). In the temperature plots (leftmost column), 
                     the heights are above the photosphere defined by $\tau_{500}$=1 layer in the solar atmosphere. }
         \label{model_examp}
   \end{figure*}

\section{Comparison of the observed and synthetic spectra and model determination}

As mentioned above, for all grids of 243 models of EBs, we could calculate the synthetic spectrum in H$\alpha$ and CaII~H
lines, which were observed by DOT. In Figure 4 we show an example of the atmospheric temperature structure for 
some selected models, along with the computed line profiles of H$\alpha$ and CaII~H. Whole grids of the line 
profiles are then used to constrain the models against our DOT observations. 
As mentioned in Sect. 2, the DOT data only contains the total emission integrated through 
the H$\alpha$ and CaII~H filters. Therefore, to make comparison with observations, it was necessary 
to transform our theoretical line profiles to the same quantity. 

We multiplied our H$\alpha$ and CaII~H computed line profiles by corresponding 
filter transmittance curves. The H$\alpha$ observations were done in the H$\alpha$ line center 
and in the line wings at $-0.7$ and $+0.7$~{\AA} from the line core. For the CaII~H line the 
observations were carried out in the line core and in the line wing at $-2.35$~{\AA}
(so that the EB signal is not contaminated by enhanced emission in the hydrogen H$\epsilon$ line). From the 
description of the DOT data we also know the FWHM of both filters: 0.25~{\AA} for H$\alpha$, 
and 1.35~{\AA} for CaII~H. We constructed normalized Gaussian functions with corresponding 
FWHM and multiplied the theoretical H$\alpha$ and CaII~H line profiles by the corresponding 
transmittance curves.  We repeated this procedure for the line profiles of all 243 models 
of our grids and for all wavelength positions of DOT filters, i.e. $-0.7$~{\AA}, line center, 
$+0.7$~{\AA} for the H$\alpha$ filter, line center, and $-2.35$~{\AA} for thew CaII~H filter.
Finally we integrated the transmitted intensities over the wavelength range of filters. These are then
used to compute the contrast according to Eq. (1).

   \begin{figure}
            \includegraphics[width=\hsize]{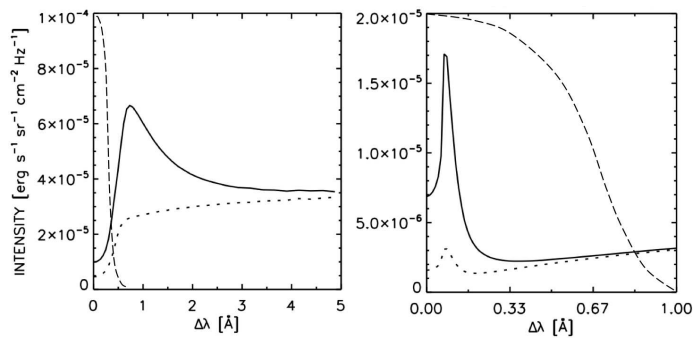}
      \caption{Comparison of filter bandpasses with the width of H$\alpha$ (left) and CaII~H lines (right).
                      Solid lines show representative line profiles of EBs, dotted lines 
                      represent the quiet-Sun line profiles, and dashed lines are the relative transmitance 
                      profiles of the DOT filters. }
         \label{filter_bandpass}
   \end{figure}

Figure 5 presents an example of theoretical H$\alpha$ and CaII~H line profiles, 
together with the Gaussian function representing the transmittance curves of the
H$\alpha$ (left) and CaII~H (right) filters. These plots are only done for the line-center 
positions of both filters, and they are identical 
at the $-0.7$ and $+0.7$~{\AA} positions for the H$\alpha$ line and at $-2.35$~{\AA} for CaII~H.
The passband of the H$\alpha$ filter is quite narrow compared to the spectral line 
width, but the FWHM of CaII~H filter is rather broad, covering both the line core and the 
emission peaks.

To work with contrasts, as we did for observations, we also multiply 
the theoretical reference H$\alpha$ and CaII~H line profiles corresponding to the semiempirical 
model C7 by the filter transmittance curves and integrated over the filter passbands. 
We thus obtained the theoretical contrasts for all models in all considered grids. 
The transmittance-function normalization factors cancel in Eq. (1) so that their
values are arbitrary for a given spectral filter.

As described in Sect. 3, we constructed three grids of EBs hot spot models. These grids 
differ by the gas-density inside EBs. For one grid we used the hot spot density identical to 
that of C7 model (GRID\_1RHO), twice larger (GRID\_2RHO), and four times larger
(GRID\_4RHO). For each grid we analyzed both the H$\alpha$ and CaII~H lines observed at
the line center and in the wings (i.e., four contrast values). Therefore, we constructed 
12 different curves, presenting the changes of the contrast across the model grids.
Such curves are presented in Fig. 6. Each plot shows the changes in the contrast calculated
for theoretical line profiles according to Eq. 1. The saw-like shape of curves reflects the 
specific construction of our grids: 9 different hot spot temperature enhancements at 27 
different height locations in the atmosphere - therefore, we get 27 maxima.
The plots in the lefthand column correspond to the H$\alpha$ line, and the righthand column is for CaII~H. 
The upper row shows the results for grid GRID\_1RHO, middle for GRID\_2RHO, and 
lower for GRID\_4RHO. Each plot contains curves of the theoretical contrast for the line centers, 
and for the line wings.

We also constructed one additional grid of 243 hot spot models of EBs with the same density 
as in C7 model but with three times higher microturbulent velocity in the hot spot region. This grid 
allowed us to analyze the influence of the microturbulence on the emergent line profiles. 
In each plot of Fig. 6 we also included  the observed contrast for EB\_1 at the time of its
maximum brightness ($C$=0.3 and 2.6, respectively) taken in the H$\alpha$ and CaII~H line wings 
($C$=1.6 and 1.5, respectively).

   \begin{figure*}
   \centering
   \mbox{
            \includegraphics[width=5cm,angle=90]{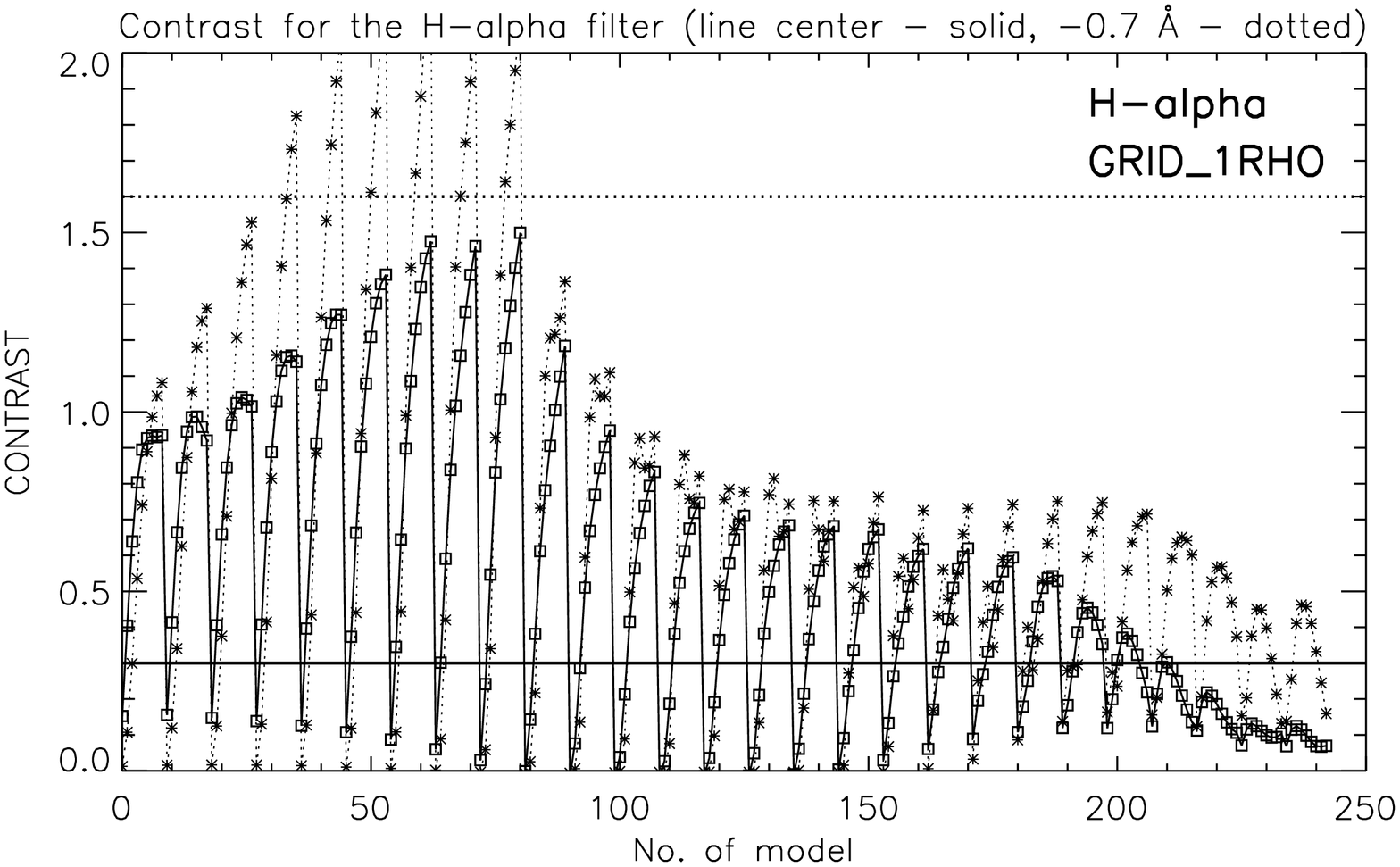}
            \includegraphics[width=5cm,angle=90]{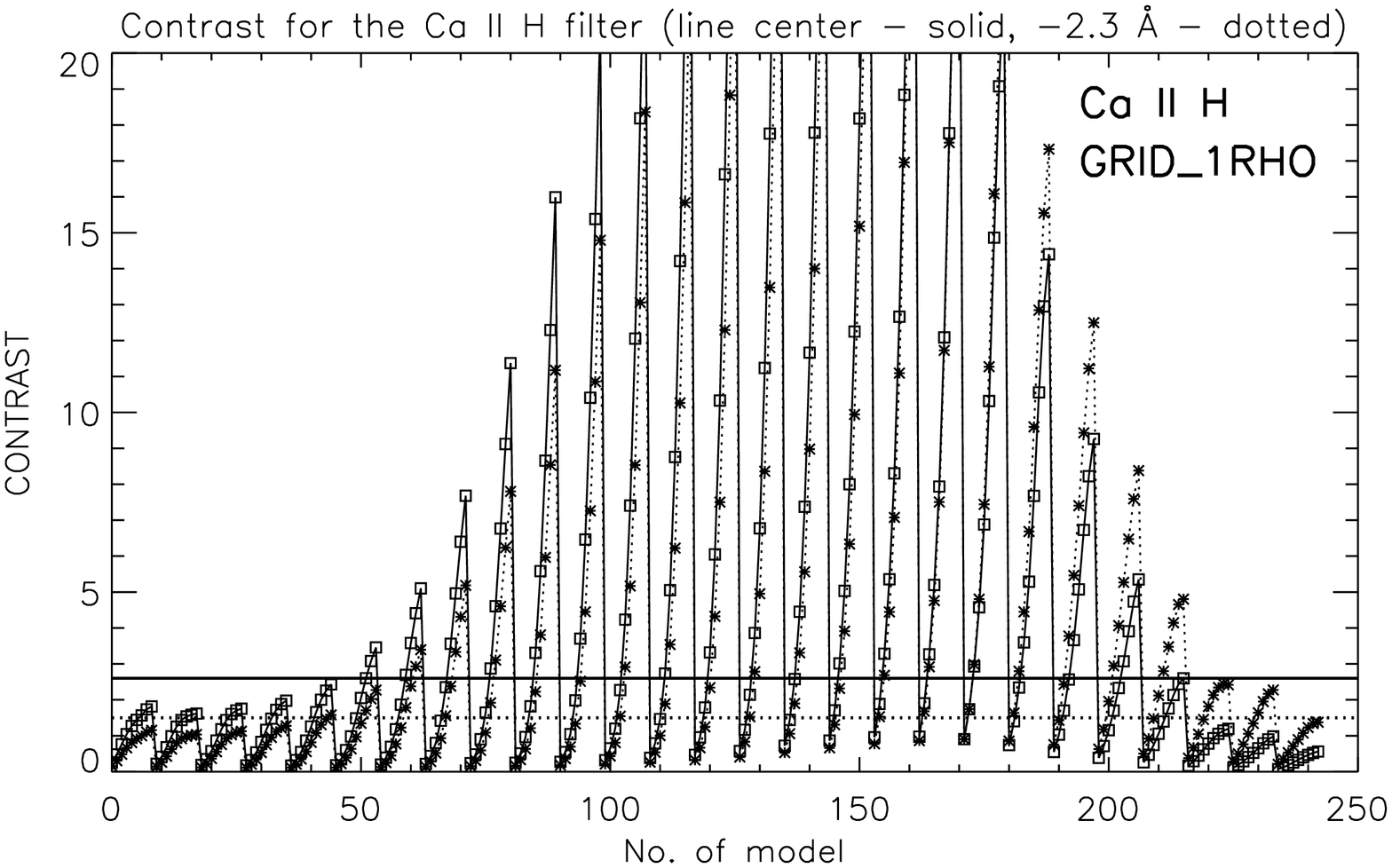}
            }
   \mbox{
            \includegraphics[width=5cm,angle=90]{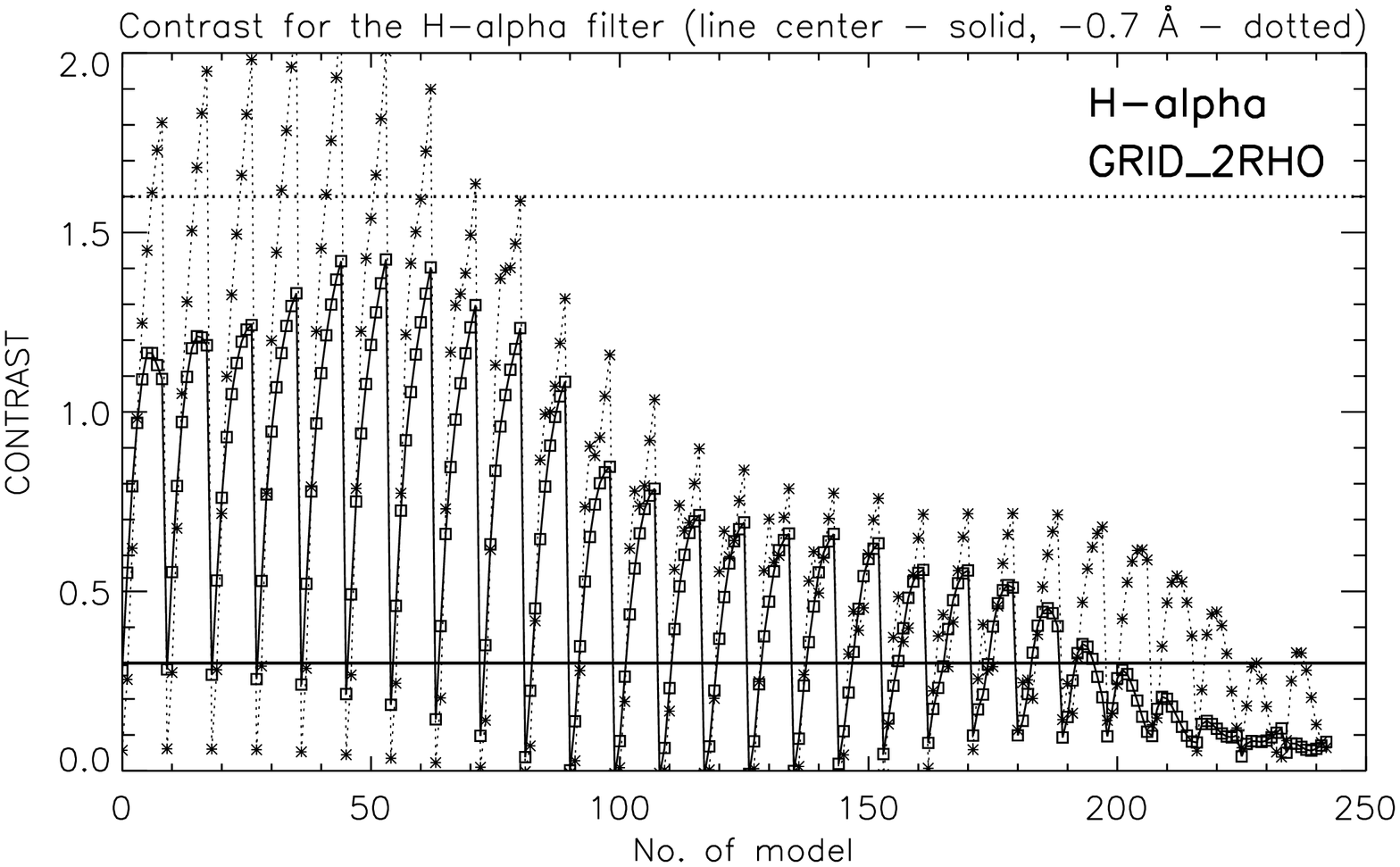}
            \includegraphics[width=5cm,angle=90]{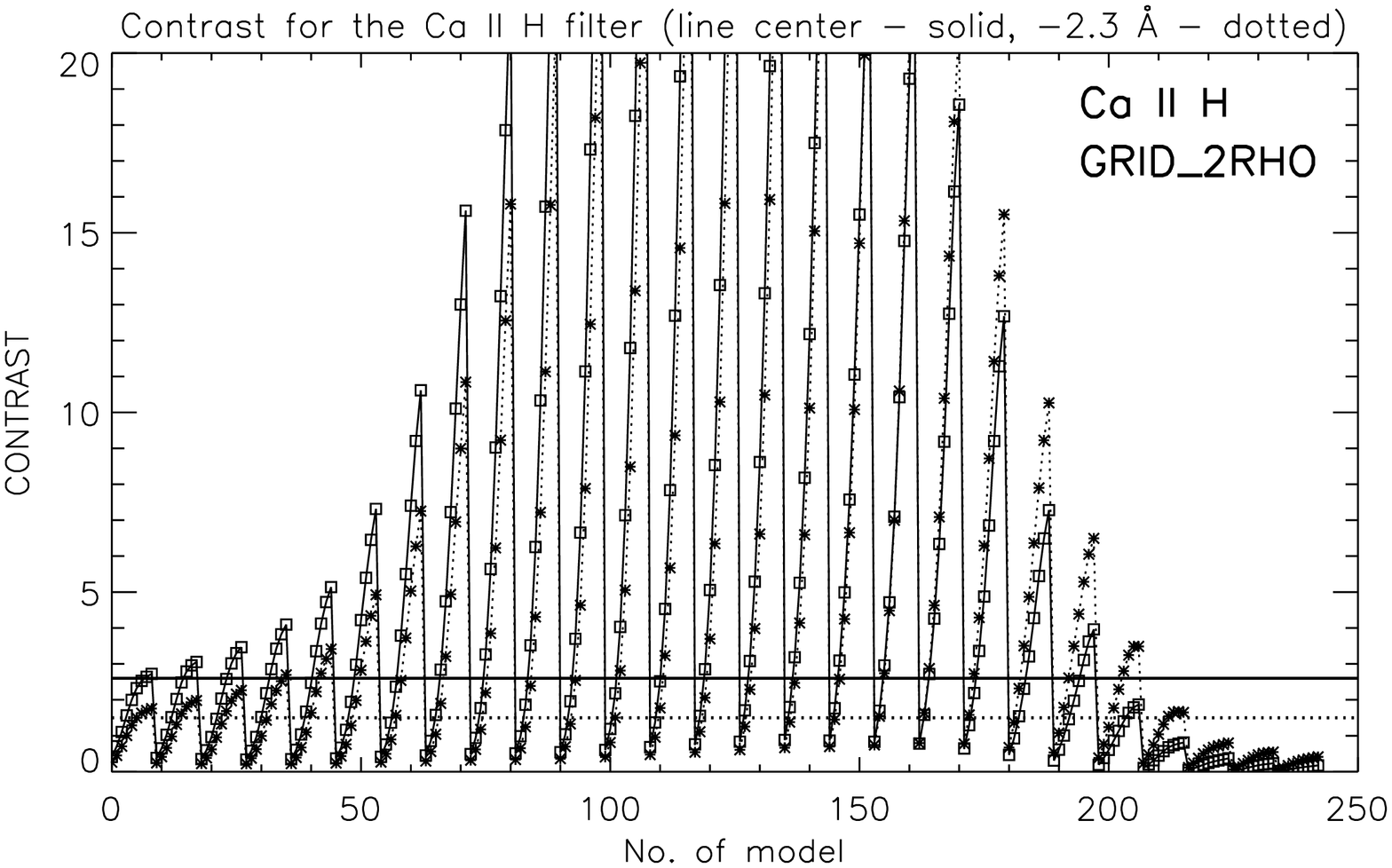}
            }
   \mbox{
            \includegraphics[width=5cm,angle=90]{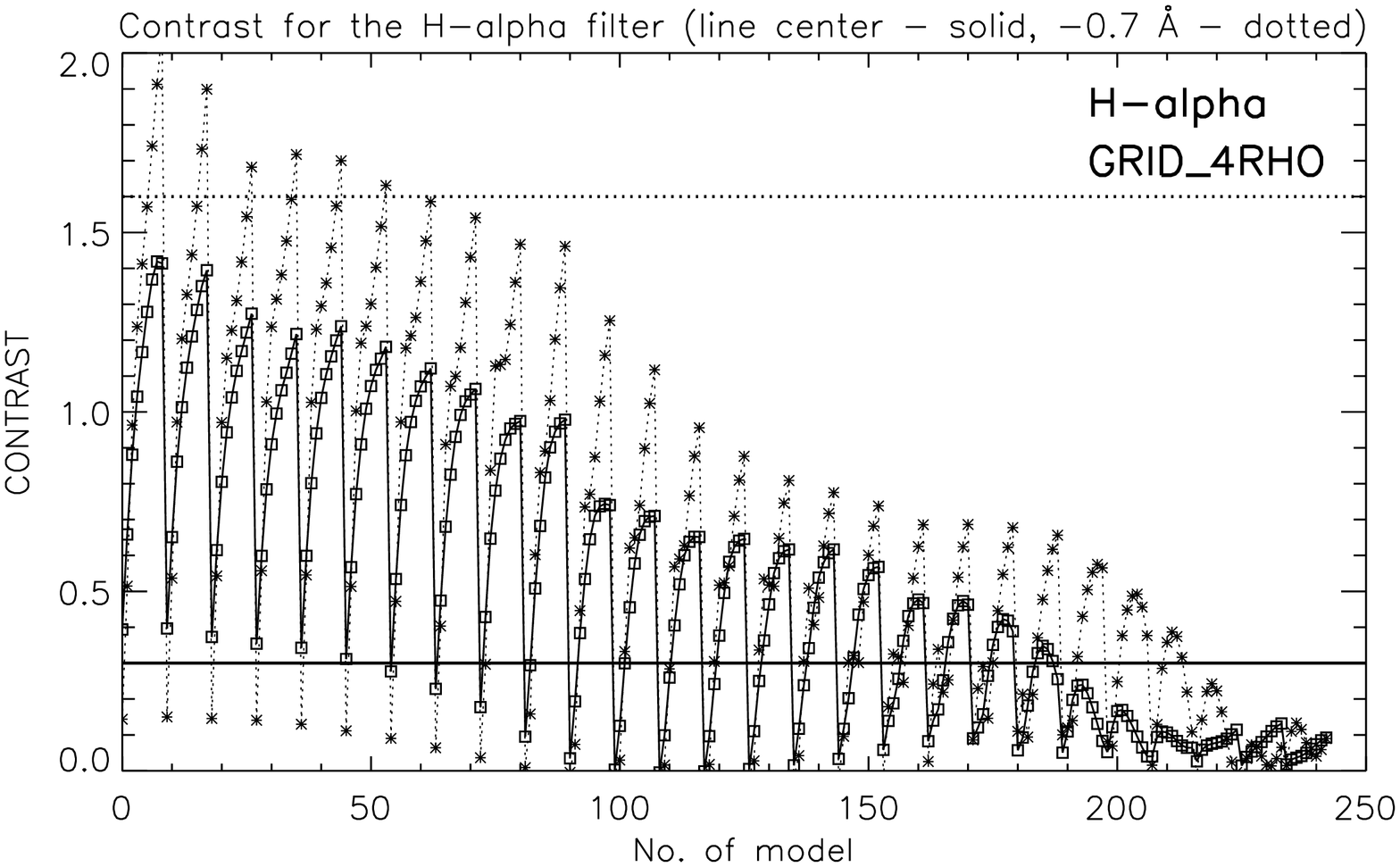}
            \includegraphics[width=5cm,angle=90]{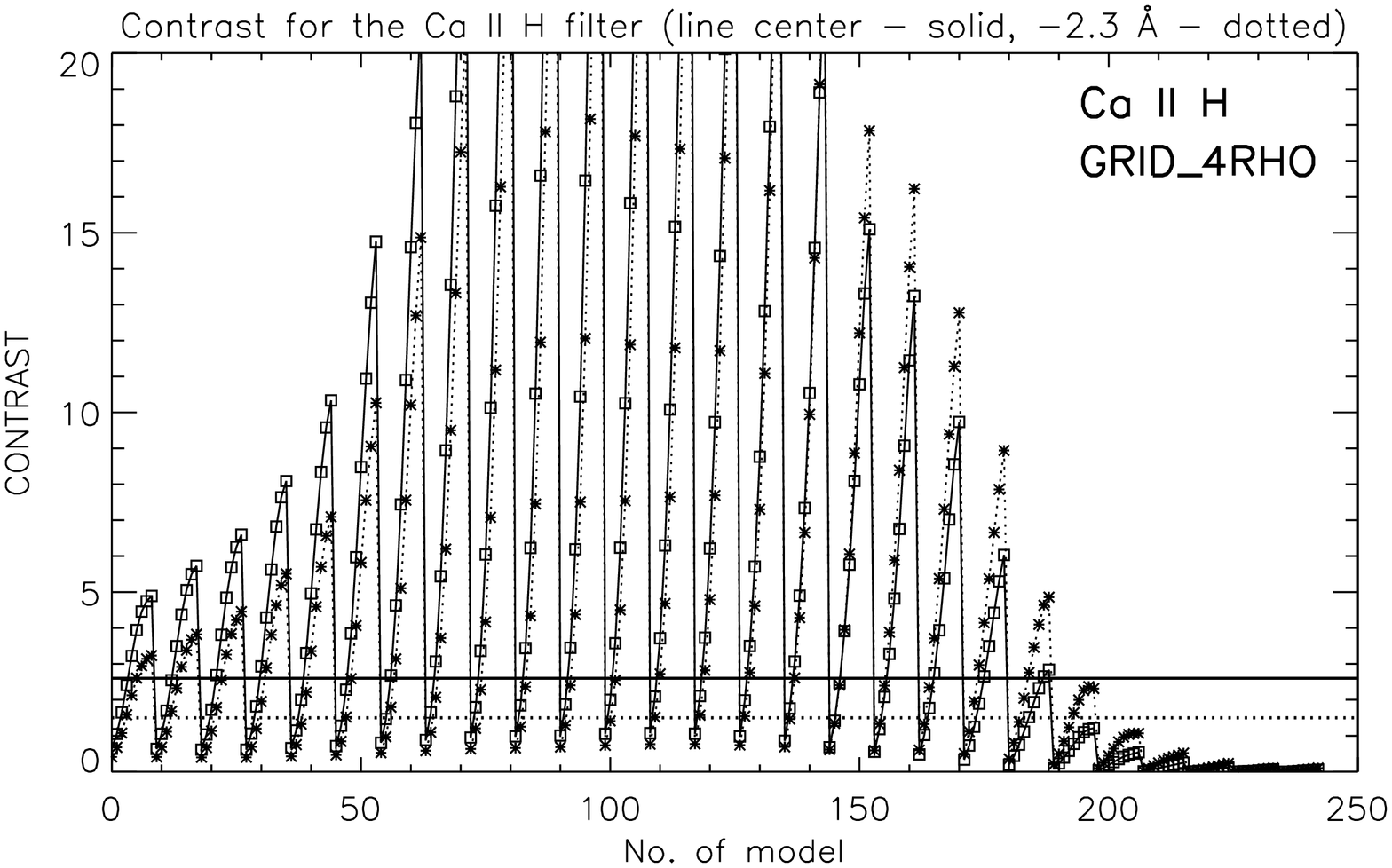}
            }               
      \caption{Variations of the theoretical contrast across the grids GRID\_1RHO 
                     (upper row), GRID\_2RHO (middle row), and GRID\_4RHO (lower row). The results 
                     for the H$\alpha$ line are presented in the lefthand column plots, for CaII~H line  
                     in the righthand column plots. In each panel there are curves calculated for the line center 
                     (solid lines with squares), and for the line wings H$\alpha~-0.7$ and CaII~H $-2.35$~{\AA} 
                     (dotted lines with stars). The horizontal lines visible in each plot represent the observed 
                     contrast for EB\_1 at the moment of its maximum brightness derived from DOT filtergrams obtained  
                     in the H$\alpha$ (plots in the left column) and in the CaII~H line (plots 
                     in the right column). Solid lines correspond to the contrast observed in lines centers, and dotted lines 
                     denote the contrast observed in line wings. }
         \label{contrast_rho}
   \end{figure*}

   \begin{figure*}
   \centering
   \mbox{
            \includegraphics[width=5cm,angle=90]{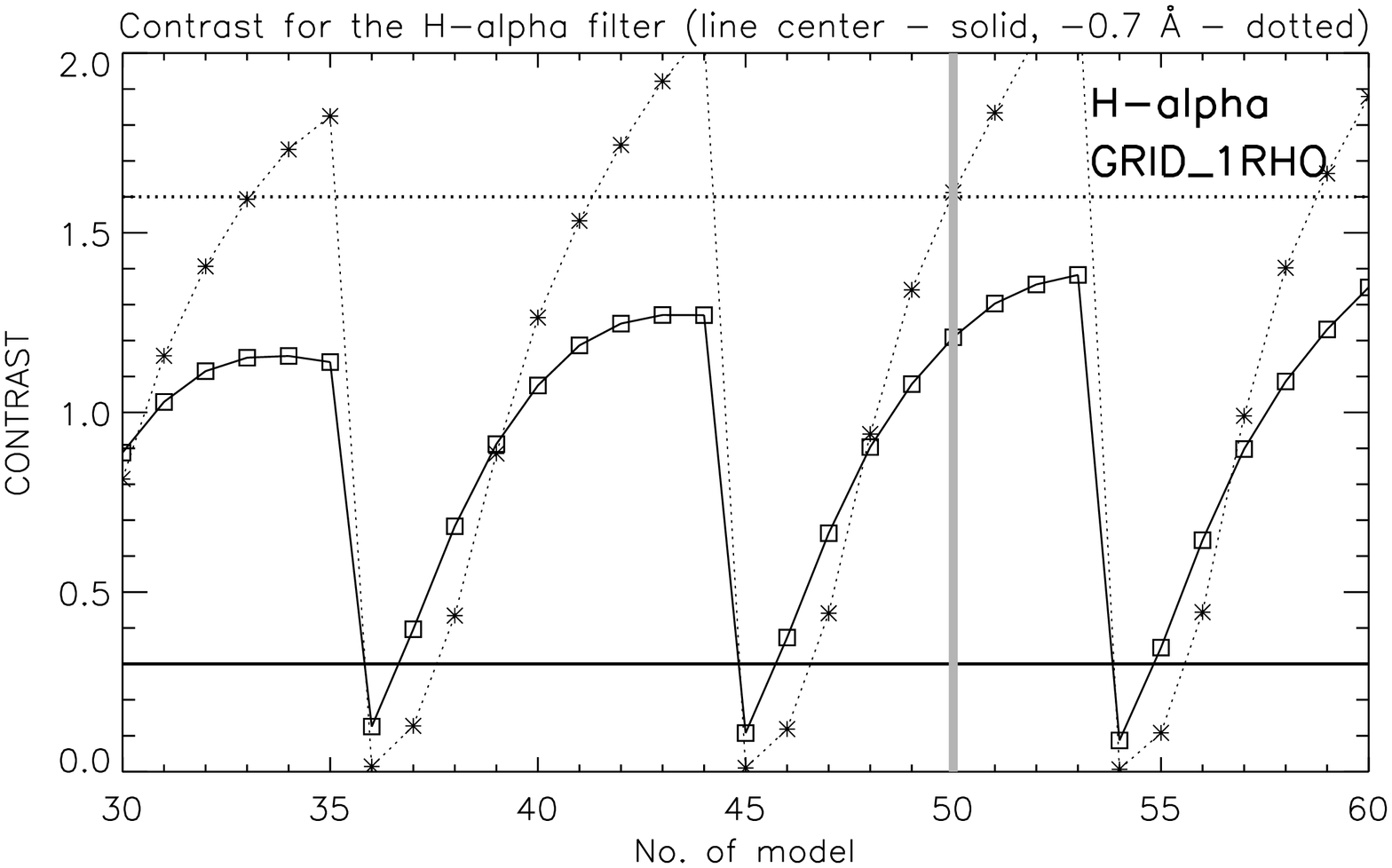}
            \includegraphics[width=5cm,angle=90]{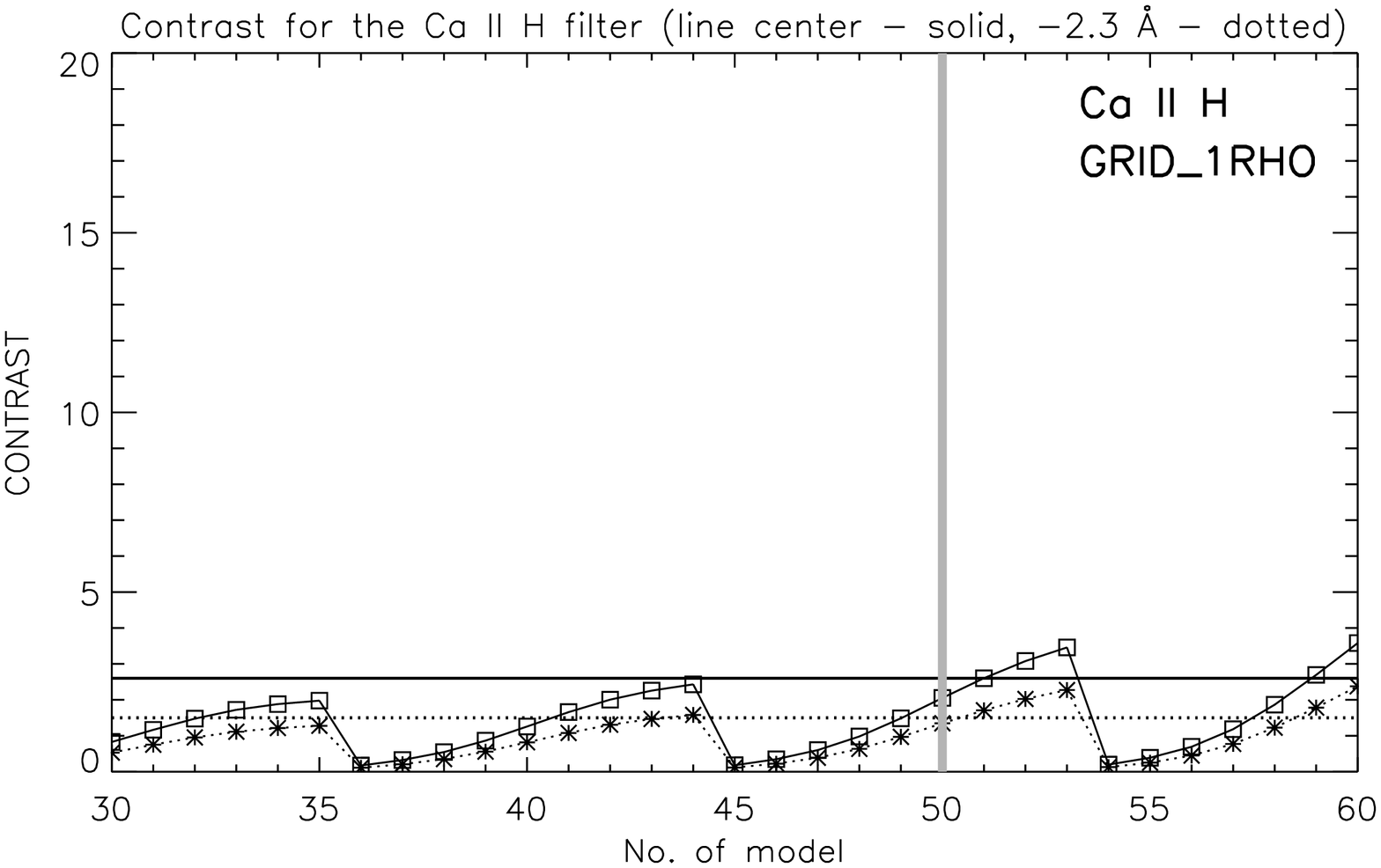}
            }
        \caption{Zoomed plots of the theoretical contrast across the grid GRID\_1RHO.
                       These plots include a small range of grid models, where the theoretical 
                       H$\alpha$ and Ca II H contrasts fit best the corresponding observed contrast. 
                       The horizontal lines represent the observed 
                       contrast for EB\_1 at the moment of its maximum brightness derived from the DOT 
                       filtergrams. Solid lines correspond to contrast observed in the lines centers, dotted lines 
                     denote the contrast observed in the line wings. Thick vertical gray lines correspond 
                     to the position of Model No. 50, which fits the observations in the best way.}
         \label{contrast_rho_details}
   \end{figure*}

From these plots it is possible to find models for which the theoretical contrast has the same (or similar) value 
as the one taken from observations. Such models correspond to positions where the horizontal 
lines (observed contrasts) cross the theoretical curves of contrasts. 
It is clearly seen that taking H$\alpha$ and CaII~H data into account separately, 
one gets many such positions where the horizontal lines (observed contrast)
cross the theoretical curves. But we are looking for one specific position
where all horizontal lines of the observed contrast cross all corresponding curves of theoretical contrasts
for the H$\alpha$ line center, H$\alpha$ - $0.7$~{\AA}, CaII~H line center, and CaII~H - $2.35$~{\AA} at the 
same position on the vertical axis, defining the model for which the theoretical contrast corresponds 
to the observed one in these two spectral lines. 

Such situation occurs for the model No. 50 of the grid GRID\_1RHO. Only for 
this model are the crossing points between the horizontal lines and the corresponding theoretical contrast 
curves for H$\alpha$ line center, H$\alpha$ - $0.7$~{\AA}, CaII~H line center, and CaII~H - $2.35$~{\AA} 
located at approximately the same position with No. 50.
That means that the theoretical emission in H$\alpha$ and CaII~H lines calculated for this model
fits the observed emission reasonably, i.e. the model No. 50 of GRID\_1RHO describes the properties
of EB\_1 well. 

Figure 7 presents the subframes (zoom) of plots from Fig. 6 for the grid with the gas-density structure 
identical to that of C7 (GRID\_1RHO). It is clearly visible that the horizontal lines, representing the observed 
contrasts of EB\_1 in the H$\alpha$ - $0.7$~{\AA}, CaII~H line center, and CaII~H - $2.35$~{\AA}, cross the 
corresponding theoretical curves in the vertical position close to model No. 50. The properties of this 
model are reflected at the location on the gray vertical line.
Only the horizontal line of the observed contrast for the H$\alpha$ line
center ($C$=0.3) does not cross the corresponding theoretical curve around model No. 50.  The closest 
model, which produces the same contrast as observed in the H$\alpha$ line center, is No. 46, 
but then this model does not fit any other observed contrasts (e.g. for H$\alpha$ - $0.7$~{\AA}).
We comment on this problem in the discussion.

In Figure 8 a, b, c we present the temperature distribution and theoretical H$\alpha$ and CaII~H line profiles 
calculated for Model No. 50 of GRID\_1RHO. These line profiles exhibit strong emission 
peaks and self-reversals. Our analysis showed that such line profiles lead to theoretical contrasts 
that are consistent with those observed in H$\alpha$ and CaII~H DOT filtergrams, with some
difference in the H$\alpha$ line core discussed in the next section.

During our study we also tested the influence of the microturbulence and partial frequency redistribution (PRD)
on the emergent synthetic line profiles. In Fig. 8 we show the CaII~H 
line profiles calculated for the same model No. 50, but using three times higher 
microturbulent velocity in the hot spot region than for C7 (plot d). 
To see the effect of PRD, 
the CaII~H line profile was calculated for model No. 50 with C7
microturbulence, but using the complete frequency
redistribution (CRD, plot e). We discuss these results in the next section.

   \begin{figure*}
   \centering
   \mbox{
            \includegraphics[width=5cm]{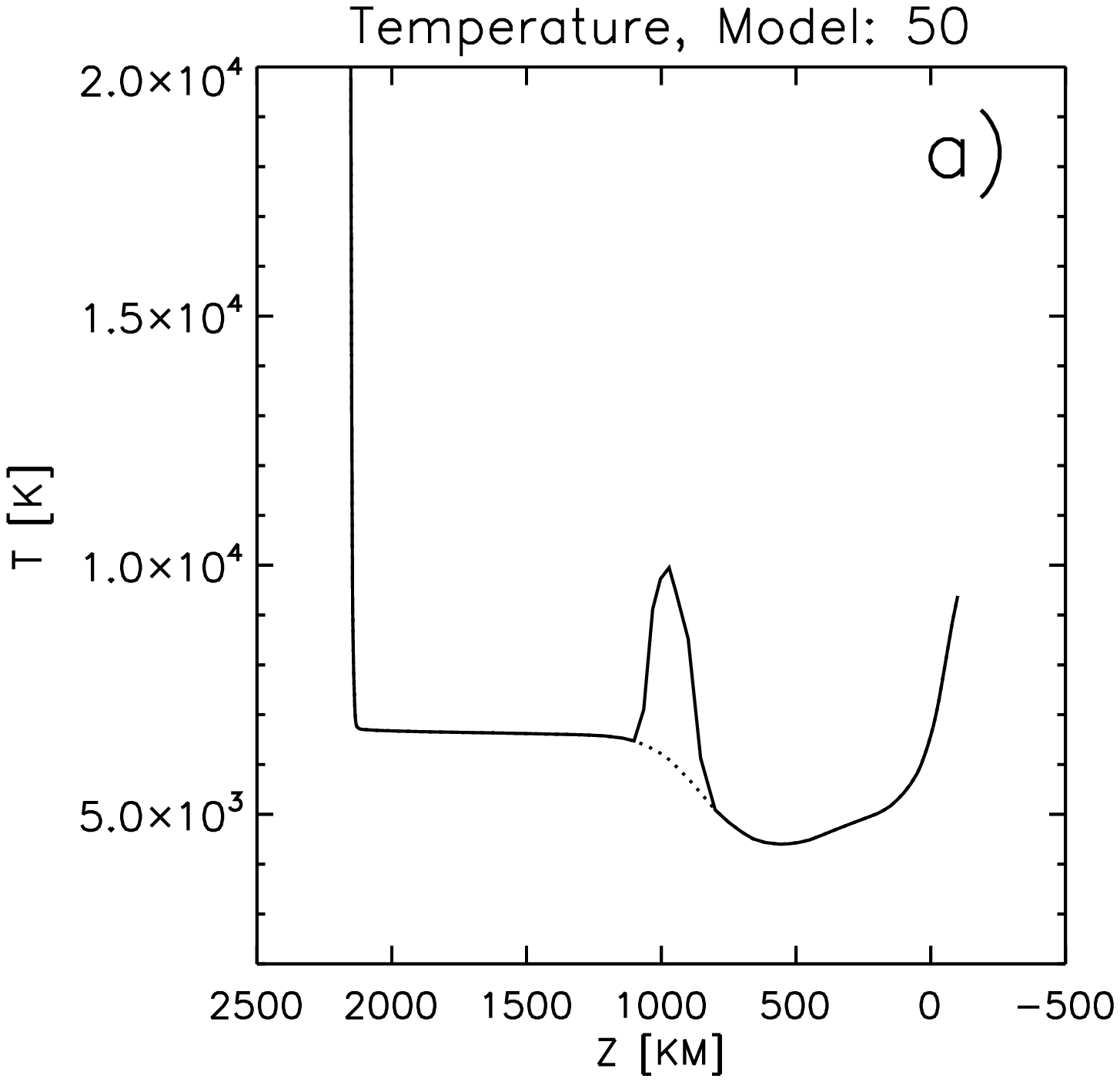}
            \includegraphics[width=5cm]{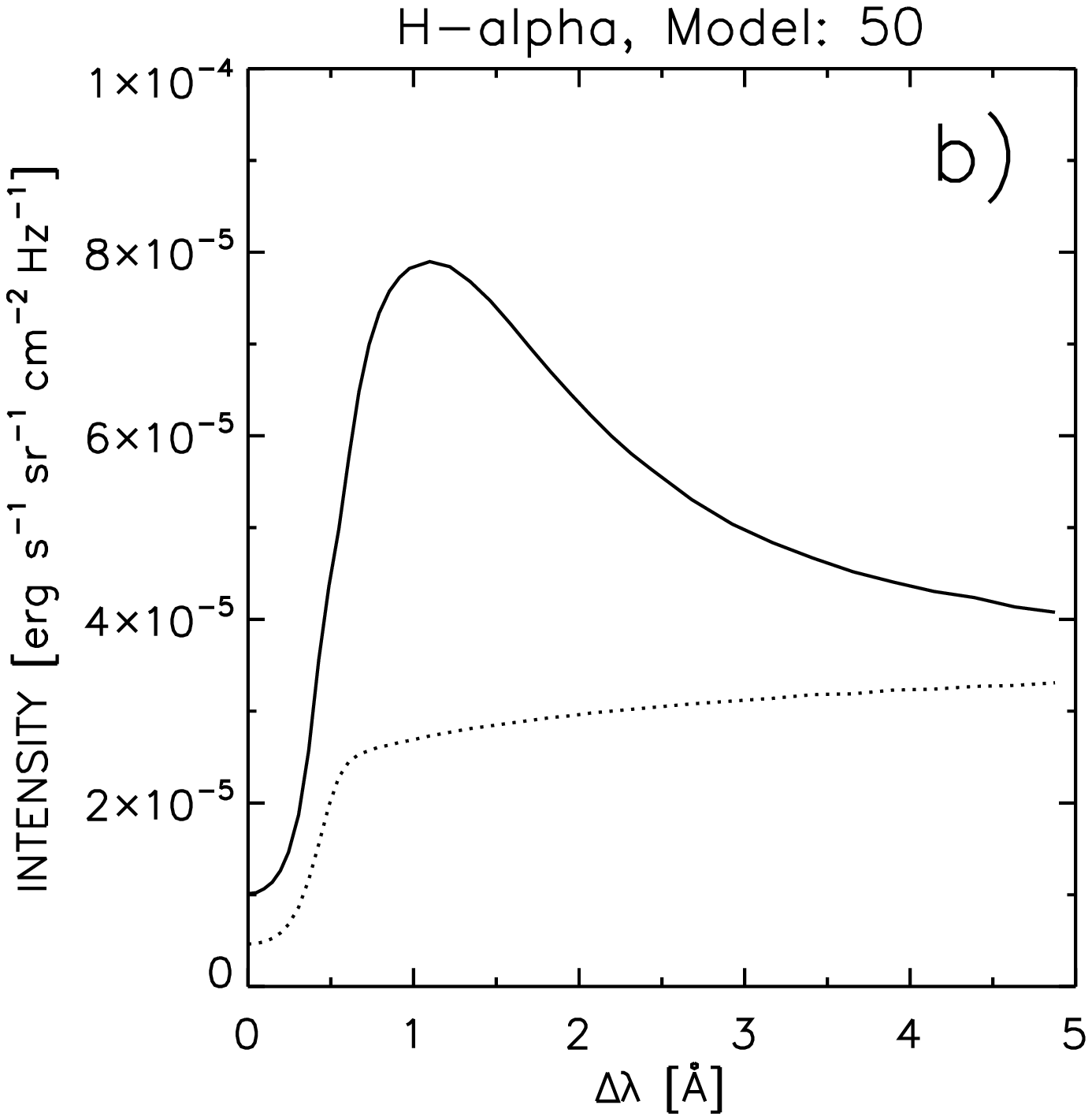}
            \includegraphics[width=5cm]{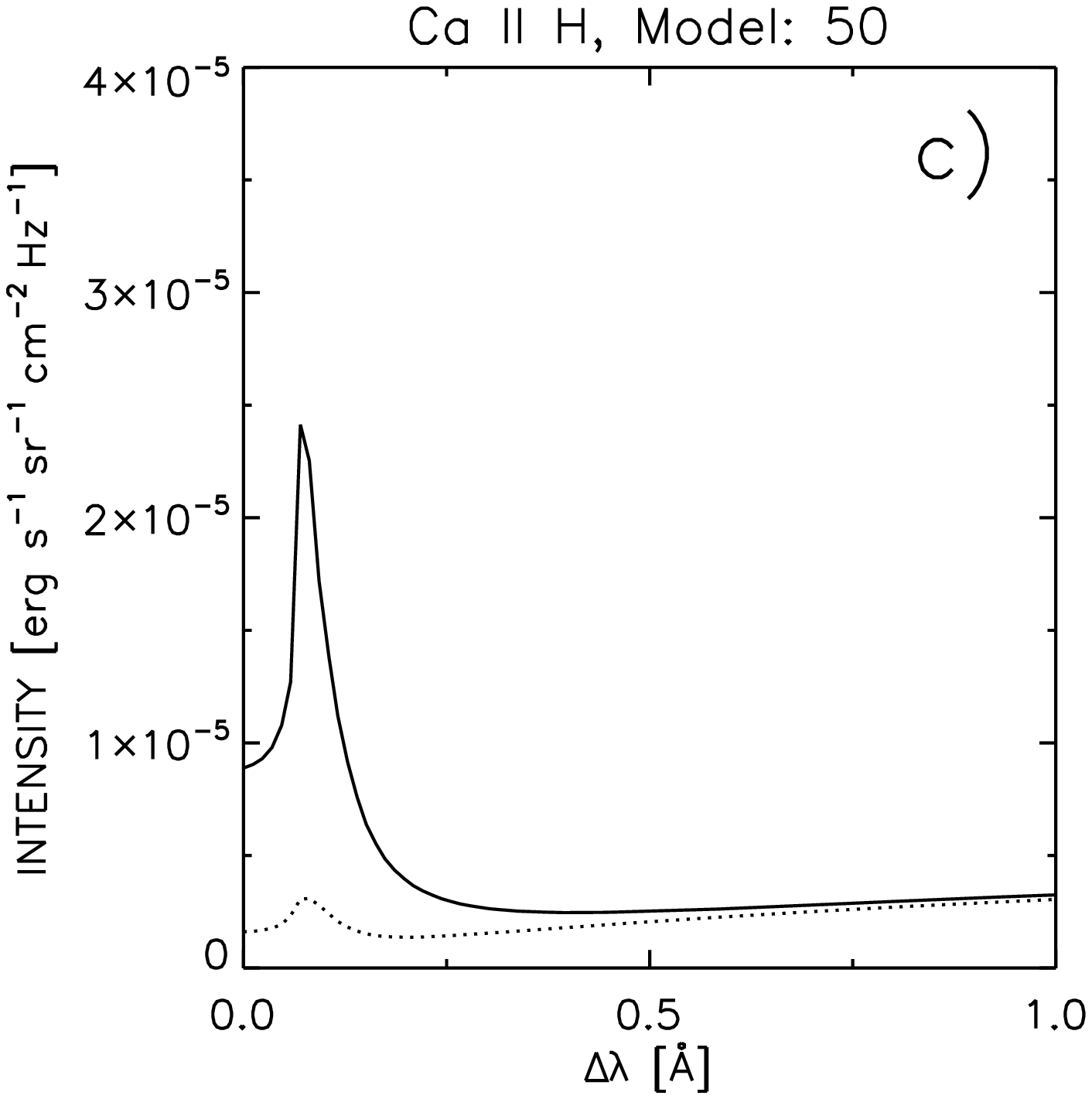}   
            }
   \mbox{
            \includegraphics[width=5cm]{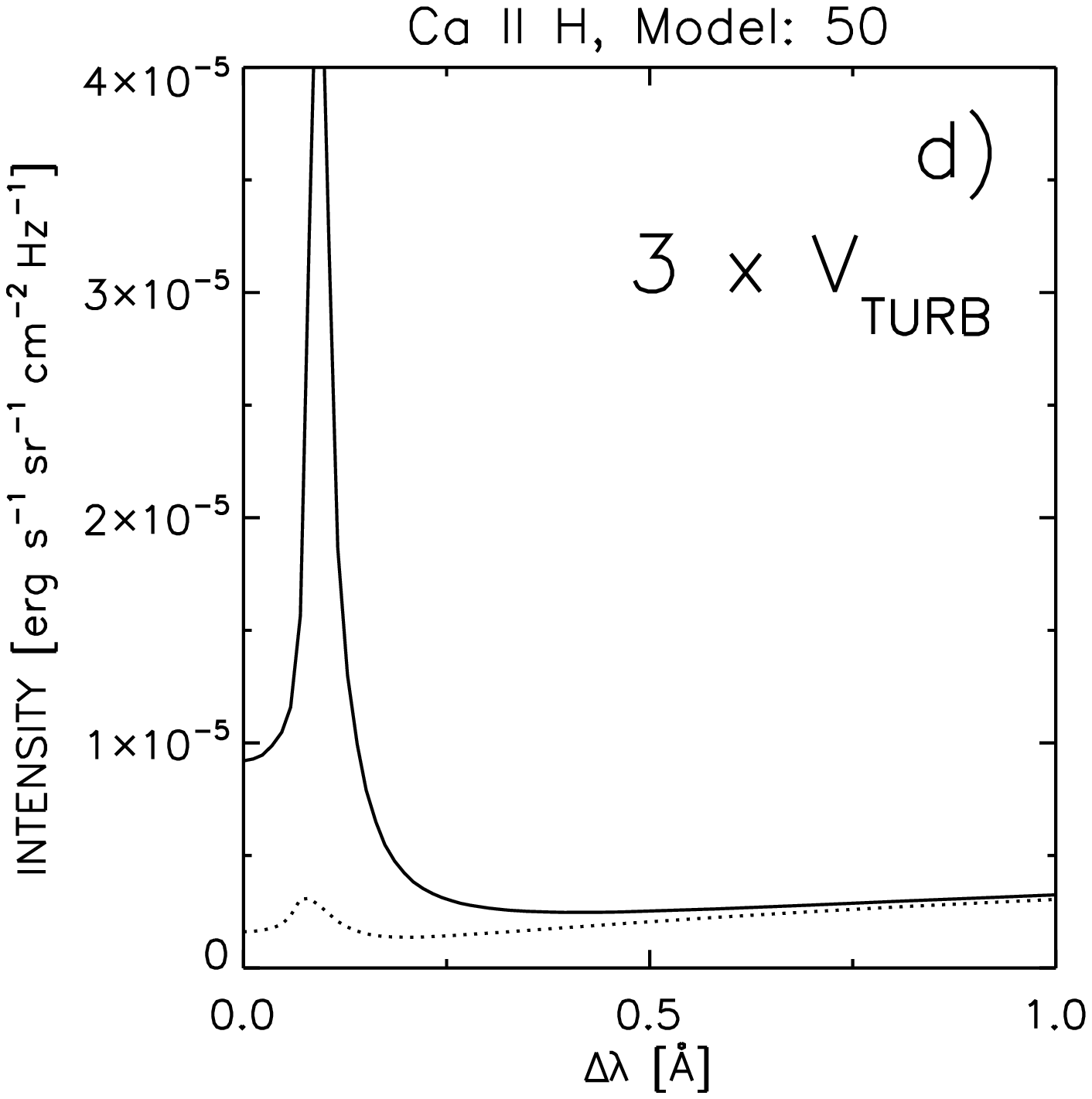}
            \includegraphics[width=5cm]{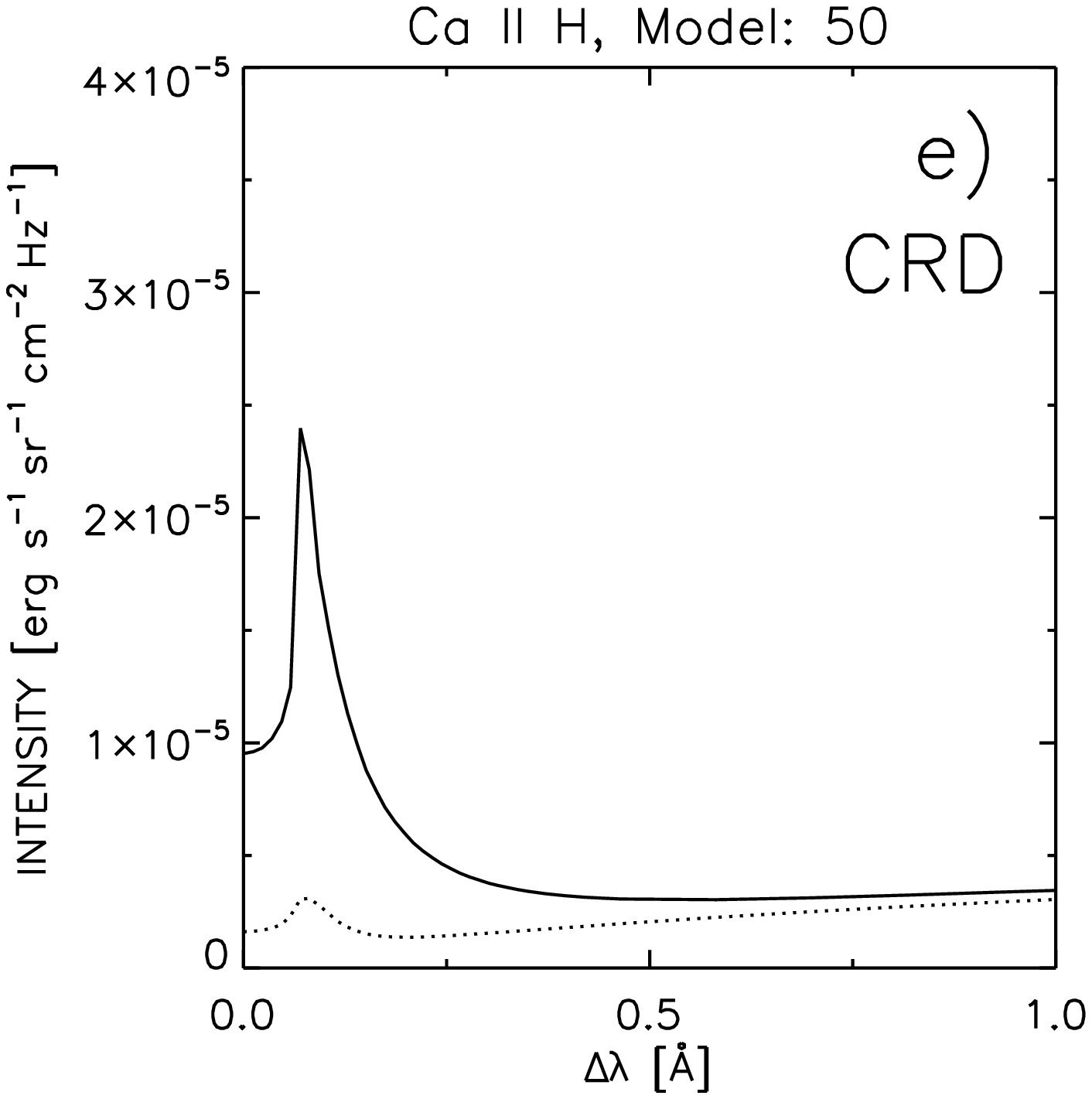}
            }
      \caption{Temperature structure and the theoretical profiles corresponding to model No. 50 
                      of the GRID\_1RHO (plots a, b, c), for which we obtained the best agreement between 
                      the theoretical and observed emission of EB\_1. Dotted curves show the temperature 
                      and line profiles corresponding to C7 model of the quiet Sun. In the lower panel we present
                      the theoretical CaII~H line profiles calculated for the same model No. 50 but with 
                       three times greater microturbulence in the hot spot region (plot d) and with the
                       complete frequency redistribution (CRD, plot e) for standard microturulence taken from C7.   
                       }
         \label{final_model}
   \end{figure*}

\section{Discussion}

\subsection{C7 model and canopy structure}

The most recent semi-empirical model of the solar atmosphere C7 
(Avrett \& Loeser 2008) is the model in 
hydrostatic equilibrium, and its temperature structure was constructed in order to
obtain an optimal fit to mean UV/EUV spectra of the quiet Sun at the disk center.
The spectra were recorded by the SUMER spectrometer onboard the SOHO satellite.
Transition-region temperatures obey the energy-balance condition, which includes
the ambipolar diffusion. However, since the model is 1D plane-parallel, the ambipolar
diffusion takes place only along the vertical direction and thus the magnetic
field must also be implicitly vertical. On the other hand, dark H$\alpha$ fibrils, which we see
in the active region where our EBs are located, suggest that the magnetic field is more inclined,
almost horizontal.  

Moreover, these fibrils forming a canopy structure certainly reach greater 
heights than the upper chromosphere of the C7 model, which is placed around
2000 km. These fibrils obviously obscure the EBs emission in the H$\alpha$ 
line center but they are not included in our semi-empirical hot spot models of EBs.
Therefore, the theoretical H$\alpha$ line profiles calculated with our NLTE codes showed 
significant emission in the line core, and the theoretical contrast $C\approx1.1$ 
is stronger than the observed one ($C$=0.3, see Fig. 7).

For a more accurate analysis, which will be based on better spectral resolution
(either high-dispersion spectra or narrow-band filtergrams like IBIS or CRISP),
we suggest deriving the fibril properties using the standard cloud model technique
(see Tziotziou 2007). With more spectral points along the line profiles, one can derive
at least three fibril parameters, i.e. the line-center optical thickness $\tau_0$, the
source function, and the Doppler width. From the line asymmetry the velocity in the
fibril can also be obtained. The idea is to take the synthetic H$\alpha$ profile (from C7
or other relevant atmosphere) as the background radiation entering the cloud from below.
The cloud model parameters can then be obtained from fibril observations close to
the position of EBs. In this way one can account for the absorption of the EB radiation
coming from lower atmospheric layers (where the hot spot is likely to be located) by the overlying
fibrils frequently present in active regions, as suggested by Rutten et al. (2013). This is
critical in the line core, while in the wings the fibrils are supposed to be transparent.   

\subsection{PRD and effect of microturbulence}

In our calculations we used a standard PRD for hydrogen Lyman $\alpha$ and $\beta$ lines.
The higher members of the Lyman series are less and less sensitive to effects of the partial coherence.
This has some influence on te resulting ionization structure and affects the Lyman $\alpha$ and
$\beta$ line profiles. On the other hand, the H$\alpha$ line is treated with complete frequency
redistribution (CRD), which is a good approximation for such subordinate line. However, PRD was 
shown to be critically important for CaII resonance lines H and K (Uitenbroek 1989), so
we use it in our modeling. In EBs, the main difference between CRD and PRD takes place
in the line wings as we show in Fig. 8. Owing to partially-coherent scattering, the line wings
of CaII H are lower than for the CRD case, and this affects our diagnostics.

The calcium lines are also known to be sensitive to microturbulent broadening. Significant
differences in CaII H and K line profiles computed with a VAL-3C microturbulent structure (Vernazza et
al. 1981) and zero microturbulence have been demonstrated in Uitenbroek (1989). For C7 model we
obtained similar results. However, when the microturbulent velocity is enhanced in the region of
the hot spot, the line peaks are surprisingly higher. We tested the case of
three times higher velocities, and the resulting profile for model No. 50 is shown in Fig. 8.
This effect will deserve more systematic study in the future because the enhanced turbulence 
in EBs (possibly caused by the reconnection) is likely to be present.

\subsection{Energetics of EBs}

Based on our semi-empirical modeling, we are able to estimate the total radiative-energy loss at the site of the
hot spot. For the best-fit model No. 50 we computed the net radiative losses for hydrogen, CaII, and MgII.
The hot spot losses peak around 1000 km and amount approximately to 100 erg sec$^{-1}$ cm$^{-3}$. 
Unlike to the quiet Sun
models where the losses are dominated by CaII and MgII (at such heights), we found the dominant contribution 
from hydrogen. At the border heights of the hot spot, they sharply decrease and go even to low negative
values, indicating possible radiative heating of atmospheric regions attached to the hot spot. The FWHM of
radiation loss function is roughly 100 km in height (corresponding to peak temperature enhancement)
so that the height-integrated losses are roughly
equal to 5 $\times$ 10$^{8}$ erg sec$^{-1}$ cm$^{-2}$. We can also compute a total volumetric energy loss 
over the EB\_1 life time on the order of 10 minutes, as observed with DOT (Fig. 3), and using the 
observed area of around 0.3 arcsec$^{2}$ (circular diameter of approximately 0.6{$\arcsec$}). 
With these numbers we finally get the total energy loss of 5~$\times$~10$^{26}$ ergs. 
Nelson et al. (2013) used their statistical analysis to estimate this energy to 10$^{22}$ - 10$^{25}$
ergs, which is lower than our result and also two to four orders of magnitude lower than the  
estimate of Georgoulis et al. (2002). However, the life time and the area of EBs obtained by Nelson et al. 
(2013) seems to be quite small compared to previous results and to our observations.
Georgoulis et al. (2002) and Fang et al. (2006) claim that each EB has a total life time energy of around 
10$^{27}$ ergs, a value similar to microflares. Our calculations obtained for EB\_1 almost reach the latter values. 
All these estimates critically depend on the temperature and
density models of EBs, as well as on the way the radiative losses are computed and on the lifetimes
and volumes of particular EBs. Our model No. 50 has an unperturbed gas density compared to C7, but the temperature
increase is fairly significant, up to 10$^4$ K at height 1000 km. Our analysis of energetics of EBs does 
not allow us to distinguish between different sources of EBs energy. However, since the calculated energy of EB is similar to 
microflares, the most probable mechanism of heating is magnetic reconnection. This idea can be supported
by the observations, where the association between EBs and emerging flux tube occurrence is clearly 
visible (Watanabe et al. 2008). To confirm the magnetic nature of EBs energy source, it would be useful
to obtain the high-resolution observations, which directly relate the amount of magnetic flux cancellation with 
the energy of EBs. Unfortunately, it is difficult to obtain such data.

\section{Conclusions}

Being aware of the limitations of the DOT spectral data and of the 
1D plane-parallel modeling, we have explored for the first time 
the range of EBs NLTE models that are able to simultaneously reproduce the hydrogen 
H$\alpha$ and CaII~H lines. We constructed extended grids of atmospheric
models aimed at describing the thermal and density structure of the EB. 
Since no MHD models applicable to our NLTE simulations are available yet, we have
followed the approach of Kitai (1983), starting from the quiet-Sun model and
perturbing it in temperature and density - we call this EB structure a hot spot 
model in analogy to recent modeling of stellar flares (Kowalski 2013). 
The semi-empirical model C7 of Avrett \& Loeser (2008) was perturbed at different
heights, increasing the temperature to a given limit. With this we considered
three types of density perturbations and also an increase in microturbulent
velocity that has a critical impact on CaII H and K lines. As in the study
of Kitai (1983), our hot spot models are not in hydrostatic equilibrium, 
which certainly leads to generating of flows actually detected in EBs.
However, these flows have rather low velocities on the order of 10 km sec$^{-1}$, 
and in our case the DOT data do not exhibit any significant asymmetry between
red and blue wing signals in H$\alpha$. Therefore we start here with simple
static models. We also show the importance of PRD for modeling CaII~H and K lines.

Using all grids of our models and the peak intensities from the time evolution
of EB\_1, we found several interesting results. First, all our models
indicate the location of the EB in the low chromosphere above the temperature
minimum region. This is fully consistent with recent spectropolarimetric
analysis of Socas-Navarro et al. (2006), who also found a hot layer
between the upper photosphere and the lower chromosphere. It also supports
previous results of Kitai (1983) and Fang et al. (2006). The temperature
enhancement needed to explain both H$\alpha$ and CaII~H lines was found 
to be rather significant. This can be compared with the results
of the above-mentioned studies. Increasing the gas
density up to three times that of C7 at the location of EB, the hot spot
shifts upward into the low chromosphere. The effect of an increased 
microturbulence on CaII~H and K lines is fairly surprising - unlike to the case 
of the quiet-Sun atmosphere (see, e.g., Uitenbroek 1989), a localized increase
in $v_t$ within the EB region (hot spot) leads to an {\em increase} in CaII~H and
K line peaks, rather than decrease.

Based on purely morphological observations at different lines and
continua, Rutten et al. (2013) claim that "EBs are the most spectacular solar
photosphere phenomenon". They explicitly conclude that EBs are not
chromospheric. Our results and others mentions above indicate that the location
of the hot spot is rather higher, in the low chromosphere above the
temperature minimum region; for example Watanabe et at. (2008) explains EBs as events 
caused by magnetic reconnection just in lower chromosphere. Other works also support 
such location of EBs (Georgoulis et al. 2002, Fang et al. 2006). 
However, none of these models did consider an
observed cotemporal enhancement of the G-band, which is formed at the
photospheric levels (our Fig. 3). This enhancement is probably due to the magnetic
concentration (MC) changes (Rutten et al. 2013) and not due to a deep photospheric
heating related to our hot spots.

Our analysis also showed that even the filtergrams can be used for quantitative 
analysis of the spectroscopic properties of the solar structures provided that the images have 
been well processed and the filters passbands are known.
However, the real spectra are much more useful in this analysis and give us better 
constraints, e.g., in model determination. Therefore, new multiwavelength observations 
of EBs have been acquired using both high-dispersion
spectrographs like DST on Hida Observatory (R. Kitai - private communication) or THEMIS
(B. Schmieder - private communication), as well as
with the narrow-band tunable filters (IBIS, CRISP). Unlike the DOT images, these 
data provide real spectra, so the reconstruction of the full line profiles for EBs is possible. 
There is an obvious advantage in using the full observed line profiles instead of filtergrams, 
because direct comparison with the theoretical line profiles is then possible, 
instead of comparison of intensities integrated within the filter passbands. 
Then, the EBs model determination will take the spectral lines 
shape into account and will therefore be much more precise.

We used the above-mentioned observations to search for
the presence of EBs and the work is in progress to analyze such data. We also
plan to use 2D/3D NLTE codes developed at Ondrejov Observatory to study
the structure and behavior of EBs further. This effort should then be coupled to realistic
MHD modeling of heating processes that lead to the apparition of EBs. 
We will extend our NLTE modeling to Mg~II~h and k resonance lines, which have already been observed in
EBs by IRIS and which will provide a new diagnostics tool. 

\begin{acknowledgements}
      The work of AB and PH was supported by the Grant Agency of the Czech Republic under 
      contract P209/11/2463 and by the project RVO-67985815 of the Astronomical Institute of
      the Academy of Sciences of the Czech Republic. 
      We would like to thank the Dutch Open Telescope staff for providing the data used in this
      study.
\end{acknowledgements}

\makeatletter
\def\@biblabel#1{}
\makeatother

\end{document}